\documentclass[a4paper,twocolumn,showpacs,pra,eqsecnum]{revtex4} 
\usepackage{graphicx}
\usepackage{dcolumn} 
\usepackage{bm} 
\usepackage{amsmath}
\usepackage{amsfonts} 
\usepackage{amssymb}
\usepackage{color}

\usepackage[normalem]{ulem}

\begin{document}
\title{Renormalization-group calculation of the superfluid/normal-fluid \newline 
interface of liquid $^4$He in gravity near $T_\lambda$}

\author{R. Haussmann}
\affiliation{Fachbereich Physik, Universit\"at Konstanz, D-78457 Konstanz, Germany}

\date{\today}

\begin{abstract}
The superfluid/normal-fluid interface of liquid $^4$He is investigated in gravity 
on earth where a small heat current $Q$ flows vertically upward or downward. We 
present a local space- and time-dependent renormalization-group (RG) calculation 
based on model $F$ which describes the dynamic critical effects for temperatures $T$ 
near the superfluid transition $T_\lambda$. The model-$F$ equations are rewritten in 
a dimensionless renormalized form and solved numerically as partial differential 
equations. Perturbative corrections are included for the spatially inhomogeneous 
system within a self-consistent one-loop approximation. The RG flow parameter is 
determined locally as a function of space and time by a constraint equation which 
is solved by a Newton iteration. As a result we obtain the temperature profile of 
the interface. Furthermore we calculate the average order parameter 
$\langle\psi\rangle$, the correlation length $\xi$, the specific heat $C_Q$ and 
the thermal resistivity $\rho_\mathrm{T}$ where we observe a rounding of the 
critical singularity by the gravity and the heat current. We compare the thermal 
resistivity with an experiment and find good qualitative agreement. Moreover we 
discuss our previous approach for larger heat currents and the self-organized 
critical state and show that our theory agrees with recent experiments in this 
latter regime.
\end{abstract}


\maketitle

\section{Introduction}
\label{Sec:01}
On earth in liquid $^4$He the gravity is an external force which causes a space dependent 
pressure $p=p(z)$ depending on the altitude coordinate $z$. Since the critical temperature 
of the superfluid transition $T_\lambda=T_\lambda(p)$ depends on the pressure $p$, in the 
helium the critical temperature $T_\lambda(z)=T_\lambda(p(z))$ varies with the altitude $z$. 
In leading approximation it is a linear function of the altitude
\begin{equation} 
T_\lambda(z) = T_\lambda(z_0) + (\partial T_\lambda / \partial z) \, (z-z_0)
\label{Eq:A01} 
\end{equation} 
where the gradient is determined experimentally as \cite{Al68}
$\partial T_\lambda / \partial z=+1.273\,\mathrm{\mu K/cm}$. The sign is positive 
which means that the critical temperature increases with the altitude $z$.

In thermal equilibrium, the local temperature of the helium $T(\mathbf{r},t)=T$ is 
constant with respect to any space and time variable. If in an experiment we choose 
the temperature $T=T_\lambda(z_0)$ we find an interface at $z=z_0$, which separates 
superfluid $^4$He in the upper region $z>z_0$ where $T<T_\lambda(z)$ from normal-fluid 
$^4$He in the lower region $z<z_0$ where $T>T_\lambda(z)$. This interface is the main 
concern of the present paper.

Correlation effects imply an interface which is not sharply defined but smeared out 
over a certain length scale $\xi_g$. Ginzburg and Sobyanin \cite{GS76} have calculated 
the order parameter profile $\psi(z)$ for liquid $^4$He in gravity within their $\psi$ 
theory which is a mean-field theory modified by scaling functions in order to 
incorporate the effects of critical fluctuations and the critical exponents to some 
extent. They find the characteristic length scale $\xi_g= 67\,\mathrm{\mu m}$ 
(see Fig.\ 4 and Eq.\ (3.49) in Ref.\ \onlinecite{GS76}).

A heat current $Q$ flowing from bottom to top in the direction of $z$ enhances the formation 
of the superfluid/normal-fluid interface. Heat transport phenomena imply a space-dependent 
temperature $T(z)$ with a negative gradient $\partial T / \partial z < 0$ which acts 
opposed to the positive gradient of the critical temperature $T_\lambda(z)$. Onuki 
\cite{On83,On87} has investigated the interface under a heat flow $Q$ within a dynamic 
mean-field theory modified by scaling functions. He finds that the thickness of the 
interface decreases according to $\xi_Q\sim Q^{-1/2}$ with increasing heat current $Q$.

While on earth the gravity acceleration $g=9.81\,\mathrm{m/s^2}$ is constant, the 
heat current $Q$ can be varied in the experiment. For large heat currents $Q\gtrsim Q_0$, 
the heat-current effects dominate, where on the other hand for small heat currents 
$Q\lesssim Q_0$ gravity effects dominate. The heat current which separates both regimes, 
is about $Q_0=70\,\mathrm{nW/cm^2}$. In this paper we focus on small heat currents where 
gravity is the main effect.

The critical dynamics of liquid $^4$He near the superfluid transition $T_\lambda$ is 
described by a hydrodynamic model with Gaussian fluctuating forces which is called 
model $F$ in the classification of Hohenberg and Halperin \cite{HH77}. This model has 
originally been derived by Halperin, Hohenberg, and Siggia \cite{HHS74} in order to 
describe the critical dynamics of a planar ferromagnet, which is in the same universality 
class as liquid $^4$He. The field-theoretic renormalization-group theory of model $F$ has 
been elaborated by Dohm \cite{Do85,Do91}. The specific heat and the thermal conductivity 
have been calculated up to two-loop order \cite{Do85} and compared with very accurate 
experimental data \cite{Al84,Al85}. In this way, the renormalized coupling parameters 
and some other parameters have been adjusted \cite{Do91} so that all parameters of 
model $F$ are known. Thus model $F$ is ready for application without any further adjustable 
parameters.

In this paper we present a renormalization-group (RG) calculation of the 
superfluid/normal-fluid interface based on model $F$. The calculation is technically very 
difficult and challenging for two reasons. First the Green functions and Feynman 
diagrams must be evaluated in a spatially inhomogeneous system. Secondly, the 
renormalization factors depend on space and time coordinates via the RG flow parameter 
so that the partial derivatives with respect to space and time must be replaced by 
appropriate \emph{covariant} derivatives.

The first challenge was overcome step by step in several previous papers. On the 
normal-fluid side of the interface the Green function was calculated \cite{HD91} 
for a zero order parameter $\langle\psi(z)\rangle=0$ and a linear temperature 
parameter $r_0(z) = a_0 + b_0 z$. The local thermal conductivity $\lambda_T(T,Q)$ 
and the related temperature profile $T(z)$ was calculated. On the superfluid side 
of the interface the Green function and related thermodynamic quantities were 
calculated \cite{HD92} for a plane-wave order parameter 
$\langle\psi(z)\rangle=\eta\,e^{ikz}$ and a constant temperature parameter $r_0$. 
Here a critical superfluid current was found which implies a depression of the 
superfluid transition temperature $T_\lambda(Q)<T_\lambda$ by a nonzero heat 
current $Q$.

Later the normal-fluid approach \cite{HD91} was extended beyond the interface into 
the superfluid region \cite{Ha99,Ha99a}. The calculation was made self consistent by 
a lowest-order $1/n$ expansion which is equivalent to the Hartree approximation of 
quantum many-particle physics. In this way the superfluid region could be reached 
where in the whole system the average order parameter $\langle\psi(z)\rangle=0$ is 
zero due to phase fluctuations related to the motion of vortices where however the 
condensate density $n_\mathrm{s}=\langle |\psi(z)|^2 \rangle$ and the superfluid 
current $\mathbf{J}_\mathrm{s}=\langle \mathrm{Im}[\psi^*\nabla\psi] \rangle$ are 
macroscopically large. The RG theory was applied locally using a local flow parameter 
which depends on the altitude coordinate $z$. The specific heat $C_Q$ and the thermal 
conductivity $\lambda_T$ were calculated for the whole superfluid/normal-fluid 
interface where the effects of the gravity acceleration $g$ and the heat current $Q$ 
were included. The temperature profile $T(z)$ was obtained by integrating the 
heat-transport equation $\mathbf{Q}=-\lambda_T \nabla T$.

In the superfluid region a nonzero temperature gradient $\nabla T$ was found which 
is due to a nonzero thermal resistivity induced by the motion of vortices and 
quantum turbulence. The theory was especially successful to describe the so called 
self-organized critical state, which was predicted by Onuki \cite{On87} and which 
was discovered in the experiment by Moeur \emph{et al.}\ \cite{Du97}. In this 
state the temperature gradient $\nabla T$ is equal to the gravity induced gradient 
$\nabla T_\lambda$ of \eqref{Eq:A01}, i.e.\ $\nabla T=\nabla T_\lambda$, so that 
the system is homogeneous over a large area in space.

However, for the superfluid/normal-fluid interface the self-consistent approach 
\cite{Ha99,Ha99a} works only for large heat currents 
$Q\gtrsim Q_0=70\,\mathrm{nW/cm^2}$ where the heat current $Q$ dominates over the 
effects of gravity $g$. For smaller heat currents this approach does not yield a result. 
The existence and motion of vortices is essential for phase fluctuations in order to 
have a zero average order parameter $\langle\psi(z)\rangle=0$. 

For small heat currents $Q\lesssim Q_0=70\,\mathrm{nW/cm^2}$ vortices are not present
so that the average order parameter $\langle\psi(z)\rangle$ is nonzero. In this case, 
a local calculation is not possible. Instead, the full model-$F$ equations must be solved 
as partial differential equations. Here the second challenge arises if the RG theory is 
involved. The RG flow parameter is determined locally by a constraint condition so that 
it will depend on space and time. This fact requires the definition of \emph{covariant} 
differential operators. A first step for this kind of theory was made by the author 
and Nikodem \cite{HN08}. The interface was investigated in thermal equilibrium where 
only the gravity acceleration $g$ is present but no heat current. The covariant 
derivatives were defined for the renormalized order parameter and for the renormalized 
temperature parameter. The renormalized Ginzburg-Landau equation was solved numerically 
as a boundary value problem by the multiple-shooting algorithm. Results for the 
order-parameter profile $\langle\psi(z)\rangle$, the correlation length $\xi$, and 
the heat capacity $C$ were obtained. However, the calculations \cite{HN08} were not 
finished and not published.

The present paper is devoted to continue, extend, and publish our recent calculations 
\cite{HN08}. We develop a local and time-dependent RG theory for small heat currents 
$Q\lesssim Q_0=70\,\mathrm{nW/cm^2}$ in order to fill the gap which our previous 
theory \cite{Ha99,Ha99a} has left. We solve the partial differential equations of 
model $F$ together with a local constraint condition for the RG flow parameter. We 
calculate the average order parameter profile $\langle\psi(z)\rangle$ and the 
temperature profile $T(z)$ for the superfluid/normal-fluid interface. Furthermore, 
we calculate the related thermodynamic and transport quantities, i.e.\ the specific 
heat $C_Q$ and the thermal conductivity $\lambda_T$ or thermal resistivity 
$\rho_T=1/\lambda_T$. The calculations are not restricted to a stationary state of a 
constant heat current $Q$. More generally, we solve the model-$F$ equations as 
time-dependent partial differential equations, so that time-dependent and transient 
effects can be handled like the propagation of second sound.

The paper is organized as follows. In Sec.\ \ref{Sec:02} we briefly describe model $F$, 
the reduced hydrodynamic model for the critical dynamics of liquid $^4$He near the 
superfluid transition. Furthermore, we explain the approximations that we use. In 
Sec.\ \ref{Sec:03} we develop our method in order to solve the model-$F$ equations together 
with local constraint conditions for the local RG theory. In Sec.\ \ref{Sec:04} we present 
our numerical results for the superfluid/normal-fluid interface in gravity where small 
heat currents are flowing upward or downward. We compare our results with the experiment 
of Chatto \emph{et al.}\ \cite{Du07} and find good agreement for the local thermal 
resistivity. In Sec.\ \ref{Sec:05} we compare our small-heat-current results with the 
large-heat-current results of our previous approach \cite{Ha99,Ha99a}. We discuss the 
stability of the solutions of our present and our previous approach. Finally, in 
Sec.\ \ref{Sec:06} we compare our present and our previous approach with other theories 
and recent experiments. We discuss and conclude to which extent our theory can describe 
mutual friction effects for larger heat currents due to the motion of vortices and quantum 
turbulence.

\section{Model and approximation}
\label{Sec:02}
The local thermodynamic properties of liquid $^4$He are described by the three standard 
hydrodynamic variables, the mass density $\rho(\mathbf{r},t)$, the mass-current density 
$\mathbf{j}(\mathbf{r},t)$, and the entropy density $\sigma(\mathbf{r},t)$. Since $^4$He 
becomes superfluid below the critical temperature $T_\lambda\approx 2\,K$, there exists 
an additional fourth hydrodynamic variable, the macroscopic wave function 
$\psi(\mathbf{r},t)$, which is the order parameter of the superfluid phase transition. 
The full hydrodynamic equations for superfluid $^4$He described by all these four 
variables have been derived long ago by Pitaevski \cite{Pi59}.

For the critical dynamics near $T_\lambda$ the mass density $\rho$ and the mass-current 
density $\mathbf{j}$ are irrelevant variables, because the related hydrodynamic modes,
first sound and viscosity effects, are fast. $\rho$ and $\mathbf{j}$ can be eliminated or 
integrated out, so that the remaining relevant variables for the critical slow modes near 
the transition (second sound and order parameter relaxation) are the order parameter 
$\psi$ and the entropy density $\sigma$. For these two relevant variables, the 
hydrodynamic equations are given by model $F$ \cite{HH77} and read
\begin{eqnarray}
\frac{\partial\psi}{\partial t} &=& - 2 \Gamma_0 \frac{\delta H}{\delta\psi^*}
+ i g_0\, \psi \frac{\delta H}{\delta m} + \theta_\psi \ ,
\label{Eq:B01} \\
\frac{\partial m}{\partial t} &=& \lambda_0 \nabla^2 \frac{\delta H}{\delta m}
- 2 g_0\, \mathrm{Im} \Bigl( \psi^* \frac{\delta H}{\delta\psi^*} \Bigr) 
+ \theta_m \ .
\label{Eq:B02} 
\end{eqnarray}
For convenience and historical reasons, the dimensionless entropy density is denoted 
by $m=\sigma/k_\mathrm{B}$. In the equations
\begin{eqnarray}
H &=&\  \int d^dr\, \bigl[ \textstyle{\frac{1}{2}} \tau_0 |\psi|^2
+ \textstyle{\frac{1}{2}} |\nabla\psi|^2 + \tilde{u}_0 |\psi|^4 \nonumber\\
&&\hspace{12mm} + \textstyle{\frac{1}{2}} \chi_0^{-1} m^2 + \gamma_0 m |\psi|^2 
- h_0 m \bigr]
\label{Eq:B03} 
\end{eqnarray}
is the free energy functional divided by $k_B T$. The Gaussian stochastic forces 
$\theta_\psi$ and $\theta_m$ incorporate the fluctuations. They are defined by the 
averages $\langle \theta_\psi\rangle=0$, $\langle \theta_m\rangle=0$, and by the 
correlations
\begin{eqnarray}
\langle \theta_\psi(\mathbf{r},t)\theta^*_\psi(\mathbf{r}^\prime,t^\prime)\rangle &=&
4\,\Gamma_0\, \delta(\mathbf{r}-\mathbf{r}^\prime)\,\delta(t-t^\prime) \ ,
\label{Eq:B04} \\
\langle \theta_m(\mathbf{r},t)\theta_m(\mathbf{r}^\prime,t^\prime)\rangle &=&
-2\,\lambda_0\,\nabla^2\, \delta(\mathbf{r}-\mathbf{r}^\prime)\,\delta(t-t^\prime) \ . \qquad
\label{Eq:B05} 
\end{eqnarray}
The dimension of the space $d$ is assumed to be arbitrary and continuous in the 
general calculations. However, eventually we set $d=3$ when evaluating explicit 
results for liquid $^4$He in a three-dimensional cell. For the calculations in the 
critical regime the model-$F$ equations \eqref{Eq:B01}-\eqref{Eq:B03} are treated by 
field-theoretic means, i.e.\ perturbation series expansion with respect to Feynman 
diagrams, renormalization, and the renormalization group. For example, the heat 
capacity and the thermal conductivity were evaluated up to two-loop order \cite{Do85}.

In this paper, we use an approximation following our previous work \cite{Ha99,Ha99a}. In 
many-particle physics this approximation is known as the \emph{Hartree approximation}
(see e.g.\ Ref.\ \onlinecite{FW71}). It is a self-consistent approximation including 
only a single one-loop diagram which is the tadpole diagram. Alternatively, the approximation 
is obtained by the $1/n$ expansion in leading order, where $n$ is the number of complex 
fields in a generalized model with a generalized order parameter 
$\Psi=(\psi_1,\ldots,\psi_n)$.

For the model-$F$ equations \eqref{Eq:B01} and \eqref{Eq:B02} the approximation is obtained 
by taking the nonequilibrium average $\langle\cdots\rangle$ for all terms and by 
performing appropriate factorizations of the averages of products of the fluctuating 
hydrodynamic variables $\psi$, $\psi^*$, and $m$. The factorizations are justified by 
inspection of the Feynman diagrams of the Hartree approximation which are shown in Fig.\ 2 
of Ref.\ \onlinecite{Ha99a}. We factorize the nonlinear terms according to
\begin{eqnarray}
2 \Bigl\langle \frac{\delta H}{\delta \psi^*} \Big\rangle &\approx& 
\bigl[\tau_0 - \nabla^2 + 4 \tilde{u}_0 \langle |\psi|^2 \rangle 
+ 2 \gamma_0 \langle m \rangle \bigr] \langle\psi\rangle \ , \qquad
\label{Eq:B06} \\
\Bigl\langle \psi \frac{\delta H}{\delta m} \Big\rangle &\approx&
\langle \psi \rangle\, \Bigl\langle\frac{\delta H}{\delta m} \Big\rangle \ ,
\label{Eq:B07} 
\end{eqnarray}
where
\begin{equation} 
\Bigl\langle\frac{\delta H}{\delta m} \Big\rangle = \chi_0^{-1} \langle m \rangle 
+ \gamma_0 \langle |\psi|^2 \rangle - h_0 \ .
\label{Eq:B08} 
\end{equation}
Without an approximation we obtain 
\begin{equation} 
-2\,\Bigl\langle \mathrm{Im} \Bigl( \psi^* \frac{\delta H}{\delta \psi^*} \Bigr) \Bigr\rangle 
= \nabla \langle \mathrm{Im}[ \psi^* \nabla \psi ] \rangle \ .
\label{Eq:B09} 
\end{equation}
Consequently, from \eqref{Eq:B01} and \eqref{Eq:B02} we obtain the approximate equations
\begin{eqnarray}
\frac{\partial \langle\psi\rangle}{\partial t} &=& - \Gamma_0 \bigl[ \tau_0 - \nabla^2 
+ 4 \tilde{u}_0 n_\mathrm{s} + 2 \gamma_0 \langle m \rangle \bigr] \langle\psi\rangle \nonumber\\
&&+ i g_0 \bigl[ \chi_0^{-1} \langle m \rangle + \gamma_0 n_\mathrm{s} - h_0 \bigr] 
\langle\psi\rangle \ ,
\label{Eq:B10} \\
\frac{\partial \langle m\rangle}{\partial t} &=& \lambda_0 \nabla^2 
\bigl[ \chi_0^{-1} \langle m \rangle + \gamma_0 n_\mathrm{s} - h_0 \bigr] 
+ g_0\, \nabla \mathbf{J}_\mathrm{s} \ , \qquad
\label{Eq:B11} 
\end{eqnarray}
where we define the condensate density $n_\mathrm{s}$ and the superfluid current density 
$\mathbf{J}_\mathrm{s}$ by
\begin{eqnarray}
n_\mathrm{s} &=& \langle |\psi|^2 \rangle \ ,
\label{Eq:B12} \\
\mathbf{J}_\mathrm{s} &=& \langle \mathrm{Im}[ \psi^* \nabla \psi ] \rangle \ ,
\label{Eq:B13} 
\end{eqnarray}
respectively. 

Next, for convenience and simplification of the equations we define the temperature 
parameters
\begin{eqnarray}
\Delta r_0 &=& 2 \chi_0 \gamma_0\, \Bigl\langle\frac{\delta H}{\delta m} \Big\rangle 
\nonumber\\
&=& 2 \chi_0 \gamma_0\, \bigl[ \chi_0^{-1} \langle m \rangle + \gamma_0 n_\mathrm{s} - h_0 \bigr] \ ,
\label{Eq:B14} \\
r_0 &=& \tau_0 + 2 \chi_0 \gamma_0 h_0 + \Delta r_0 \nonumber\\
&=& \tau_0 + 2 \chi_0 \gamma_0\, \bigl[ \chi_0^{-1} \langle m \rangle + \gamma_0 n_\mathrm{s}\bigr] \ ,
\label{Eq:B15}
\end{eqnarray}
and the modified temperature parameter 
\begin{equation}
r_1 = r_0 + 4 u_0 n_\mathrm{s} \ ,
\label{Eq:B16}
\end{equation}
where $u_0=\tilde{u}_0 - \frac{1}{2} \chi_0 \gamma_0^2$ is a combined coupling constant
following Ref.\ \onlinecite{Do85}. Thus, the model-$F$ equations can be written in the 
simple form
\begin{eqnarray}
\frac{\partial \langle\psi\rangle}{\partial t} &=& - \Gamma_0 \bigl[ r_1 - \nabla^2 \bigr] 
\langle\psi\rangle + i \frac{g_0}{2 \chi_0 \gamma_0} \Delta r_0 \langle\psi\rangle\ , \qquad
\label{Eq:B17} \\
\frac{\partial \langle m\rangle}{\partial t} &=& - \nabla \mathbf{q} \ .
\label{Eq:B18}
\end{eqnarray}
The last equation is the heat transport equation where 
$\langle m \rangle = \langle\sigma\rangle / k_\mathrm{B}$ is the dimensionless 
entropy density and
\begin{equation} 
\mathbf{q} = - \frac{\lambda_0}{2 \chi_0 \gamma_0} \nabla \Delta r_0 - g_0 \mathbf{J}_\mathrm{s}
\label{Eq:B19} 
\end{equation}
is the dimensionless entropy current density. The latter is related to the heat 
current $\mathbf{Q}$ in standard physical units by 
$\mathbf{q}=\mathbf{Q}/k_\mathrm{B} T\approx \mathbf{Q}/k_\mathrm{B} T_\lambda$. 

The order-parameter equation \eqref{Eq:B17} can be written in the form 
$L \langle\psi\rangle=0$ where the operator $L$ is defined in (3.13) of our previous 
paper \cite{Ha99a} and related to the off-diagonal matrix elements of the inverse 
Green function. This observation shows that the factorizations of the present 
approach are equivalent to the self-consistent approximation in our previous paper.
We note that the factorization is applied only in the order parameter equation 
\eqref{Eq:B17}. The heat transport equation \eqref{Eq:B18} is derived without 
any factorization or approximation.

The parameters $\Delta r_0$ and $r_0$ are related to the local space and time 
dependent temperature $T=T(\mathbf{r},t)$ and to the critical temperature 
$T_\lambda=T_\lambda(z)$ of \eqref{Eq:A01} according to 
\begin{eqnarray}
\Delta r_0 &=& 2 \chi_0 \gamma_0\, \Bigl\langle\frac{\delta H}{\delta m} \Big\rangle 
= 2 \chi_0 \gamma_0\, \frac{T-T_0}{T_\lambda} \ , \qquad
\label{Eq:B20} \\
r_0 - r_{0\mathrm{c}} &=& 2 \chi_0 \gamma_0\, \frac{T-T_\lambda}{T_\lambda} \ ,
\label{Eq:B21}
\end{eqnarray}
where $T_0$ is a constant reference temperature. These equations have been derived in our 
previous paper \cite{Ha99a}. The critical value of $r_0$ is $r_{0\mathrm{c}}=0$ in one-loop 
approximation \cite{Do85} and hence also in our self-consistent approximation. The factor 
$T_\lambda$ in the denominators is easily explained. Since $H$ is the free energy divided 
by $k_\mathrm{B} T_\lambda$ and since $m$ is the entropy density divided by $k_\mathrm{B}$, 
we find that the functional derivative $\delta H / \delta m$ is a temperature divided by 
$T_\lambda$. We note that the critical temperature $T_\lambda=T_\lambda(z)$ defined in 
\eqref{Eq:A01} depends on the altitude $z$. Since the gradient is very small, the $z$ 
dependence is very weak. Thus, in the denominator we may approximately use a constant 
average value which may be the critical temperature at the interface $z=z_0$, 
i.e.\ $T_\lambda=T_\lambda(z)\approx T_\lambda(z_0)$.

Until now, the condensate density $n_\mathrm{s}$ and the superfluid current density 
$\mathbf{J}_\mathrm{s}$ defined in \eqref{Eq:B12} and \eqref{Eq:B13} are unknown. Since 
they are defined by an average of two fields $\psi$ and $\psi^*$ they are related to 
the equal-time Green function 
\begin{eqnarray} 
G(\mathbf{r},t;\mathbf{r}^\prime,t) &=& \langle \psi(\mathbf{r},t) 
\psi^*(\mathbf{r}^\prime,t)\rangle \nonumber\\
&=& \langle\psi(\mathbf{r},t)\rangle\, \langle\psi^*(\mathbf{r}^\prime,t)\rangle \nonumber\\
&&+ \langle \delta\psi(\mathbf{r},t) \delta \psi^*(\mathbf{r}^\prime,t)\rangle \ .
\label{Eq:B22} 
\end{eqnarray}
This Green function was evaluated in the Appendix of Ref.\ \onlinecite{Ha99a}. However,
while in our previous paper the average order parameter $\langle\psi\rangle$ was zero,
in the present paper it is nonzero. Hence, we must split the Green function into two 
contributions, a mean-field term and a fluctuating term where 
$\delta\psi= \psi- \langle\psi\rangle$ is the fluctuating field. While the mean-field term 
is expressed in terms of the average order parameter $\langle\psi\rangle$, the fluctuating 
term is given by the result of our previous paper. Consequently, the condensate density 
$n_\mathrm{s}$ and the superfluid current density $\mathbf{J}_\mathrm{s}$ are split into 
two contributions, too. From Eq.\ (3.24) and (3.25) of Ref.\ \onlinecite{Ha99a} we obtain
\begin{eqnarray} 
n_\mathrm{s} &=& | \langle\psi\rangle |^2
- \frac{2}{\varepsilon} A_d\, \Phi_{-1+\varepsilon/2}(X)\, r_1^{1-\varepsilon/2} \ ,
\label{Eq:B23} \\
\mathbf{J}_\mathrm{s} &=& \mathrm{Im}[ \langle\psi^*\rangle \nabla \langle\psi\rangle ]
\nonumber\\
&&+ \frac{g_0}{2 \Gamma_0^\prime} \frac{\nabla \Delta r_0}{2 \chi_0 \gamma_0} \,
\frac{1}{\varepsilon} A_d\, \Bigl( 1 - \frac{\varepsilon}{2} \Bigr) 
\, \Phi_{\varepsilon/2}(X)\, r_1^{-\varepsilon/2} \ . \qquad
\label{Eq:B24} 
\end{eqnarray}
Here it is $\varepsilon=4-d$ where $d$ is the dimension of the space. Furthermore, 
$A_d=S_d\,\Gamma(1-\varepsilon/2)\Gamma(1+\varepsilon/2)$ is a geometrical factor 
which is related via $S_d=\Omega_d/(2\pi)^d$ to the surface of the $d$ dimensional 
unit sphere $\Omega_d=2\,\pi^{d/2}/\Gamma(d/2)$. The function $\Phi_\alpha(X)$ is 
defined by the divergent series
\begin{equation} 
\Phi_\alpha(X) = \sum_{N=0}^\infty \frac{\Gamma(\alpha+3N)}{\Gamma(\alpha)}
\,\frac{X^N}{N!}
\label{Eq:B25} 
\end{equation}
where the argument $X$ is related to the square of the gradients of the parameters
$r_1$ and $\Delta r_0$ according to
\begin{eqnarray} 
X &=& \frac{1}{12\, r_1^3} \Bigl[ (\nabla r_1)^2 
+ 2 \frac{\Gamma_0^{\prime\prime}}{\Gamma_0^\prime} 
\Bigl( \frac{g_0}{4 \chi_0 \gamma_0 \Gamma_0^\prime} \nabla \Delta r_0 \Bigr)
 \cdot \nabla r_1 \nonumber\\
&&- \Bigl( \frac{g_0}{4 \chi_0 \gamma_0 \Gamma_0^\prime} \nabla \Delta r_0 \Bigr)^2
\Bigr] \ .
\label{Eq:B26} 
\end{eqnarray}
The Green function \eqref{Eq:B22} was evaluated locally for a spatial inhomogeneous 
system where the temperature parameters $r_1$ and $\Delta r_0$ depend on the space 
coordinate $\mathbf{r}$. Gradient terms $\nabla r_1$ and $\nabla \Delta r_0$ are 
included but curvature terms and higher derivatives are omitted. This fact is 
clearly seen in the function \eqref{Eq:B25} and its argument \eqref{Eq:B26}.

Now, all quantities are determined. The approximate model-$F$ equations 
\eqref{Eq:B17}-\eqref{Eq:B18} together with the entropy current density \eqref{Eq:B19},
the temperature parameters \eqref{Eq:B20}, \eqref{Eq:B21}, \eqref{Eq:B16}, and the 
quantities \eqref{Eq:B23}-\eqref{Eq:B26} are closed equations, which in principle 
can be solved numerically. We insert the condensate density \eqref{Eq:B23} into the 
equation for the modified temperature parameter \eqref{Eq:B16}. After reordering 
the terms we obtain
\begin{equation} 
r_1 \Bigl\{ 1 + 8 u_0 \frac{1}{\varepsilon} A_d\, 
\Phi_{-1+\varepsilon/2}(X)\, r_1^{-\varepsilon/2} \Bigr\}
= r_0 + 4 u_0 | \langle\psi\rangle |^2 \ .
\label{Eq:B27} 
\end{equation}
The left-hand side shows clearly that this is an implicit equation for the parameter 
$r_1$. Furthermore, we insert the superfluid current \eqref{Eq:B24} into the formula 
for the entropy current \eqref{Eq:B19}. After  reordering the terms we obtain
\begin{eqnarray} 
\mathbf{q} &=& - \frac{\lambda_0}{2 \chi_0 \gamma_0} \Bigl\{ 1 + 
\frac{g_0^2}{2 \lambda_0 \Gamma_0^\prime} 
\frac{1}{\varepsilon} A_d\, \Bigl( 1 - \frac{\varepsilon}{2} \Bigr) 
\, \Phi_{\varepsilon/2}(X)\, r_1^{-\varepsilon/2} \Bigr\} \nonumber\\
&&\times \nabla \Delta r_0 
- g_0 \mathrm{Im}[ \langle\psi^*\rangle \nabla \langle\psi\rangle ] \ .
\label{Eq:B28} 
\end{eqnarray}
Eqs.\ \eqref{Eq:B27} and \eqref{Eq:B28} of the present paper should be compared 
with Eqs.\ (3.32) and (3.35) of our previous paper \cite{Ha99a}, respectively. 
New contributions are those terms on the right hand sides which involve the 
average order parameter $\langle\psi\rangle$. The last term in \eqref{Eq:B28}
may be interpreted as the mean-field contribution of the superfluid current.
The fluctuating term of the superfluid current \eqref{Eq:B24} is proportional to 
the temperature-parameter gradient $\nabla\Delta r_0$. For this reason, the 
fluctuating term is integrated into the first term of \eqref{Eq:B28} and hence
contributes to the normal-fluid term. Similarly, in Eq.\ \eqref{Eq:B27} the 
mean-field contribution of the condensate density is put on the right-hand side 
while the fluctuating contribution is put on the left-hand side of the equation.

\section{Local renormalization-group theory for partial differential equations}
\label{Sec:03}
The liquid $^4$He is considered in the critical regime for temperatures $T$ close to 
the superfluid transition at $T_\lambda$. In order to treat the critical fluctuations 
correctly, we must renormalize the equations of the previous section and apply the 
renormalization-group (RG) theory. Since we consider local physical quantities which 
are functions of space and time, the RG flow parameter will be local and depend on 
space and time. The derivatives with respect to space and time in the model-$F$ 
equations are in conflict with a local RG flow parameter because they do not commute 
with this parameter. For this reason, the development of the local RG theory for the 
model-$F$ equations which are partial differential equations is a very challenging task.

\subsection{Renormalization}
\label{Sec:03A}
We start with the renormalization of the average order parameter $\langle\psi\rangle$, 
the temperature parameters $\Delta r_0$, $r_0$, and the coupling constant $u_0$. 
Following Ref.\ \onlinecite{Do85} we have
\begin{eqnarray}
\langle\psi\rangle &=& Z_\phi^{1/2} \langle\psi_\mathrm{ren}\rangle \ ,
\label{Eq:C01} \\
\Delta r_0 &=& Z_r\, \Delta r \ ,
\label{Eq:C02} \\
r_0 - r_{0\mathrm{c}} &=& Z_r\, r \ ,
\label{Eq:C03} \\
u_0 &=& Z_u Z_\phi^{-2} (\mu^\varepsilon/A_d) \, u \ .
\label{Eq:C04}
\end{eqnarray}
In these and the following renormalization equations we use the convention that the 
bare quantities are always on the left-hand side while renormalized quantities are 
always on the right hand side. The $Z$ factors are the renormalization factors. 
In the Hartree approximation, which we use in the present paper and in our previous 
paper \cite{Ha99a}, these $Z$ factors are
\begin{equation}
Z_\phi = 1 \ , \qquad Z_r = Z_u = 1 / [ 1 - 8u / \varepsilon] \ ,
\label{Eq:C05}
\end{equation}
where it is $r_{0\mathrm{c}}=0$. The modified temperature parameter $r_1$ is not 
renormalized. We apply the renormalizations to Eq.\ \eqref{Eq:B27}, multiply both 
sides with the inverse factor $Z_r^{-1}$, and reorder the terms. Without any further 
approximation we obtain
\begin{eqnarray} 
&&r_1 \Bigl\{ 1 + \frac{8 u}{\varepsilon} \Bigl[ \Phi_{-1+\varepsilon/2}(X)\, 
\Bigl( \frac{r_1}{\mu^2} \Bigr)^{-\varepsilon/2} - 1 \Bigr] \Bigr\} \nonumber\\
&&= r + 4 u \frac{\mu^\varepsilon}{A_d} | \langle\psi\rangle |^2 \ .
\label{Eq:C06} 
\end{eqnarray}

The average entropy density $\langle m \rangle$, the entropy current density 
$\mathbf{q}$, and the remaining model-$F$ parameters are renormalized by \cite{Do85}
\begin{eqnarray}
\langle m \rangle &=& (\chi_0 Z_m)^{1/2}\, \langle m_\mathrm{ren} \rangle \ ,
\label{Eq:C07} \\
\mathbf{q} &=& (\chi_0 Z_m)^{1/2}\, \mathbf{q}_\mathrm{ren} \ ,
\label{Eq:C08} \\
\chi_0 \gamma_0 &=& (\chi_0 Z_m)^{1/2} Z_r (\mu^\varepsilon/A_d)^{1/2}\, \gamma \ ,
\label{Eq:C09} \\
g_0 &=& (\chi_0 Z_m)^{1/2} (\mu^\varepsilon/A_d)^{1/2}\, g \ ,
\label{Eq:C10} \\
\lambda_0 / \chi_0 &=& Z_\lambda^{-1}\, \lambda \ ,
\label{Eq:C11} \\
\Gamma_0 &=& Z_\Gamma^{-1}\, \Gamma \ .
\label{Eq:C12}
\end{eqnarray}
The dimensionless renormalized parameters are defined by the ratios
\begin{eqnarray}
w &=& \Gamma / \lambda \ ,
\label{Eq:C13} \\
F &=& g / \lambda \ ,
\label{Eq:C14} \\
f &=& F^2/w^\prime = g^2 / \lambda \Gamma^\prime \ .
\label{Eq:C15}
\end{eqnarray}
We note that $\Gamma=\Gamma^\prime + i \Gamma^{\prime\prime}$ and 
$w=w^\prime + i w^{\prime\prime}$ are complex parameters. The $Z$ factors, which 
we need explicitly in our calculation, are given in Hartree approximation 
\cite{Ha99a} by
\begin{equation}
Z_m Z_\lambda = 1 / [ 1 - f / 2 \varepsilon] \ , \qquad Z_\Gamma = 1 \ .
\label{Eq:C16}
\end{equation}
The factor $\chi_0 Z_m$ will cancel out in all our equations. Hence this 
latter factor is not needed explicitly. Applying the renormalizations to 
Eq.\ \eqref{Eq:B28} we obtain the renormalized heat current 
\begin{eqnarray} 
\mathbf{q}_\mathrm{ren} &=& - \frac{\lambda}{2 \gamma}
\Bigl( \frac{A_d}{\mu^\varepsilon} \Bigr)^{1/2} \Bigl\{ 1 + \frac{f}{2 \varepsilon} 
\Bigl[ \Bigl( 1 - \frac{\varepsilon}{2} \Bigr) \nonumber\\
&&\times \Phi_{\varepsilon/2}(X) \Bigl( \frac{r_1}{\mu^2} \Bigr)^{-\varepsilon/2}
- 1 \Bigr] \Bigr\} \, \nabla \Delta r \nonumber\\
&&- g\, \Bigl( \frac{\mu^\varepsilon}{A_d} \Bigr)^{1/2} 
\mathrm{Im}[ \langle\psi_\mathrm{ren}^*\rangle \nabla \langle\psi_\mathrm{ren}\rangle ] \ .
\label{Eq:C17} 
\end{eqnarray}
Again no further approximation is made when reordering the terms. In order to 
evaluate the function $\Phi_\alpha(X)$ we need the argument $X$ expressed in terms 
of the dimensionless renormalized parameters. From \eqref{Eq:B26} we obtain
\begin{eqnarray} 
X &=& \frac{1}{12\, r_1^3} \Bigl[ (\nabla r_1)^2 
+ 2 \frac{w^{\prime\prime}}{w^\prime} 
\Bigl( \frac{F}{4 \gamma w^\prime} \nabla \Delta r \Bigr) \cdot \nabla r_1 \nonumber\\
&&- \Bigl( \frac{F}{4 \gamma w^\prime} \nabla \Delta r \Bigr)^2
\Bigr] \ .
\label{Eq:C18} 
\end{eqnarray}
The renormalization of the model-$F$ equations is straight forward. From \eqref{Eq:B17} 
and \eqref{Eq:B18} we obtain
\begin{eqnarray}
\frac{\partial \langle\psi_\mathrm{ren}\rangle}{\partial t} &=& 
- \Gamma \bigl[ r_1 - \nabla^2 \bigr] \langle\psi_\mathrm{ren}\rangle 
+ i \frac{g}{2 \gamma} \Delta r \langle\psi_\mathrm{ren}\rangle\ , \qquad
\label{Eq:C19} \\
\frac{\partial \langle m_\mathrm{ren} \rangle}{\partial t} &=& - \nabla 
\mathbf{q}_\mathrm{ren} \ .
\label{Eq:C20}
\end{eqnarray}
We furthermore need a relation between the entropy density $\langle m_\mathrm{ren} \rangle$ 
and the temperature parameters $r$ or $\Delta r$ in renormalized form. We solve 
Eq.\ \eqref{Eq:B15} with respect to the entropy density $\langle m\rangle$, eliminate the 
condensate density $n_\mathrm{s}$ by \eqref{Eq:B16}, and then perform the renormalization.
As a result we obtain
\begin{equation}
\langle m_\mathrm{ren} \rangle = m_\mathrm{c,ren}
+ \Bigl( \frac{A_d}{\mu^\varepsilon} \Bigr)^{1/2} \frac{r}{2 \gamma} 
\Bigl\{ 1 + \frac{\gamma^2}{2u} \Bigl[ 1 - \frac{r_1}{r} \Bigr] \Bigr\} \ . 
\label{Eq:C21}
\end{equation}
We have separated the constant value $m_\mathrm{c,ren}$ which is the entropy at 
the critical point with temperature $T=T_\lambda$, zero heat current 
$\mathbf{Q}=\mathbf{0}$ and zero gravity $g=0$. We need not know this constant value 
explicitly. Another useful quantity is the derivative of $\langle m_\mathrm{ren} \rangle$ 
with respect to the temperature parameter $r$. It is related to the renormalized specific 
heat \cite{Do85} according to
\begin{equation}
C_\mathrm{ren} = 2 \gamma \Bigl( \frac{\mu^\varepsilon}{A_d} \Bigr)^{1/2} 
\, \frac{\partial \langle m_\mathrm{ren} \rangle}{\partial r} = 
1 + \frac{\gamma^2}{2u} \Bigl[ 1 - \frac{\partial r_1}{\partial r} \Bigr] \ . 
\label{Eq:C22}
\end{equation}
In this way, the time derivative of the renormalized entropy density can be expressed 
in terms of a time derivative of a temperature parameter. We find
\begin{equation}
\frac{\partial \langle m_\mathrm{ren} \rangle}{\partial t} 
= \frac{C_\mathrm{ren}}{2\gamma} \Bigl( \frac{A_d}{\mu^\varepsilon} \Bigr)^{1/2}
\, \frac{\partial r}{\partial t}
= \frac{C_\mathrm{ren}}{2\gamma} \Bigl( \frac{A_d}{\mu^\varepsilon} \Bigr)^{1/2}
\, \frac{\partial \Delta r}{\partial t} \ . 
\label{Eq:C23}
\end{equation}
Since the critical temperature $T_\lambda(z)$ does not depend on the time, the two 
temperature parameters $r$ and $\Delta r$ differ by a time-independent value. For 
this reason, the time derivatives of $r$ and $\Delta r$ are equal. In the present 
paper we prefer the latter time derivative. In this way, we reformulate the second 
model-$F$ equation \eqref{Eq:C20} as
\begin{equation}
\frac{C_\mathrm{ren}}{2\gamma} \Bigl( \frac{A_d}{\mu^\varepsilon} \Bigr)^{1/2}
\, \frac{\partial \Delta r}{\partial t} 
= - \nabla \mathbf{q}_\mathrm{ren} \ .
\label{Eq:C24}
\end{equation}
In the renormalized specific heat \eqref{Eq:C22} the remaining derivative 
$\partial r_1 / \partial r$ may be obtained as the proportionality factor of the 
gradients $\nabla r_1$ and $\nabla r$ according to 
\begin{equation}
\nabla r_1 = \frac{\partial r_1}{\partial r} \, \nabla r \ .
\label{Eq:C25}
\end{equation}
In order to find a relation between the two gradients we apply the nabla operator 
$\nabla$ to Eq.\ \eqref{Eq:C06}. Thus, we find
\begin{eqnarray} 
&&(\nabla r_1) \Bigl\{ 1 + \frac{8 u}{\varepsilon} \Bigl[ 
\Bigl( 1 - \frac{\varepsilon}{2} \Bigr) \Phi_{\varepsilon/2}(X)
\, \Bigl( \frac{r_1}{\mu^2} \Bigr)^{-\varepsilon/2} - 1 \Bigr] \Bigr\} \nonumber\\
&&= \nabla r + 4 u \frac{\mu^\varepsilon}{A_d} 
\, \nabla | \langle\psi_\mathrm{ren}\rangle |^2 \ .
\label{Eq:C26} 
\end{eqnarray}
In this result the derivative has increased the index $\alpha$ of the function 
$\Phi_\alpha(X)$ by one. Furthermore, the function is multiplied by a factor 
$(1 - \varepsilon/2)$. These facts are well known from the calculations in our 
previous paper \cite{Ha99a}. By comparing Eqs.\ \eqref{Eq:C25} and \eqref{Eq:C26} 
we extract $\partial r_1 / \partial r$. Since we consider the space dependence only 
in one dimension $z$ which is the altitude, we obtain a unique result. We conclude 
that in this subsection we have derived all equations in renormalized form which 
are needed for a numerical calculation to solve the model-$F$ equations as partial 
differential equations with respect to space and time.

\subsection{Dimensionless renormalized quantities}
\label{Sec:03B}
In the renormalization equations \eqref{Eq:C01}-\eqref{Eq:C04} and 
\eqref{Eq:C07}-\eqref{Eq:C12} the new arbitrary parameter $\mu$ occurs which 
has the unit of an inverse length scale. Consequently, this parameter may be used 
to fix the length scale. On the other hand in the renormalized model-$F$ equations 
\eqref{Eq:C19} and \eqref{Eq:C24} together with \eqref{Eq:C17} the dynamic parameters 
$\Gamma=\Gamma^\prime+i\Gamma^{\prime\prime}$, $\lambda$, and $g$ all have the 
unit of a diffusion constant, i.e.\ length square divided by time. Hence, these 
parameters multiplied by $\mu^2$ may be used to fix the time scale. The 
dimensionless ratios \eqref{Eq:C13}-\eqref{Eq:C15} imply that only one of these 
parameters is needed. Thus, we will use $g \mu^2$ to fix the time scale.

We rewrite the renormalized model-$F$ equations and the related renormalized variables 
and parameters in a dimensionless form using $\mu$ and $g \mu^2$ for the scales. 
Following our previous paper \cite{Ha99a} we define the dimensionless temperature 
parameters
\begin{eqnarray}
\Delta\rho &=& \Delta r / \mu^2 = \tau^{-1} \, (T-T_0)/T_\lambda \ ,
\label{Eq:C27} \\
\rho &=& r / \mu^2 = \tau^{-1} \, (T-T_\lambda)/T_\lambda \ ,
\label{Eq:C28} \\
\rho_1 &=& r_1 / \mu^2 \ .
\label{Eq:C29}
\end{eqnarray}
The last equality sign in \eqref{Eq:C27} and \eqref{Eq:C28} is obtained by 
renormalizing the bare equations \eqref{Eq:B20} and \eqref{Eq:B21}. The renormalization 
factors are combined into the dimensionless parameter 
\begin{equation}
\tau = \Bigl( \frac{A_d \mu^d}{\chi_0 Z_m} \Bigr)^{1/2} \frac{1}{2\gamma}
\label{Eq:C30}
\end{equation}
which may be viewed as a renormalization-group (RG) flow parameter. 
A change of the length scale by replacing $\mu\to \mu\ell$ causes a change of $\tau$. 
While $\ell$ is the conventional RG flow parameter related to the length scale,
$\tau=\tau(\ell)$ is a RG flow parameter related to the temperature scale. In the 
literature on the dynamic RG theory for liquid $^4$He both flow parameters have been 
used \cite{Do85,Do91,Ha99,Ha99a,Ha99b}. For the dimensionless coupling parameters of 
model $F$ the notations $u(\ell)$, $\gamma(\ell)$, \emph{etc.}\ and $u[\tau]$, $\gamma[\tau]$, 
\emph{etc.}\ have been used. In the present paper we will use $\tau$ as the RG flow
parameter.

We define the dimensionless renormalized order parameter $Y$ and the dimensionless 
renormalized heat current $\tilde{\mathbf{q}}$ by
\begin{eqnarray}
Y &=& \langle \psi_\mathrm{ren} \rangle / \mu^{(d-2)/2} \ ,
\label{Eq:C31} \\
\tilde{\mathbf{q}} &=& \Bigl( \frac{A_d}{\mu^\varepsilon} \Bigr )^{1/2} 
\frac{\mathbf{q}_\mathrm{ren}}{g} \, \frac{1}{\mu^{d-1}}
= \frac{\mathbf{q}}{g_0} \, \frac{1}{\mu^{d-1}} \ ,
\label{Eq:C32}
\end{eqnarray}
respectively. For convenience of the notation, following Ref.\ \onlinecite{Ha99a} 
we define the dimensionless amplitudes
\begin{eqnarray}
A &=& \varepsilon^{-1}[ \Phi_{-1+\varepsilon/2}(X)\, \rho_1^{-\varepsilon/2} - 1 ] \ ,
\label{Eq:C33} \\
A_1 &=& \varepsilon^{-1}[ (1 - \varepsilon/2) \Phi_{\varepsilon/2}(X)\, 
\rho_1^{-\varepsilon/2} - 1 ] \ .
\label{Eq:C34}
\end{eqnarray}
Consequently, the renormalized heat current \eqref{Eq:C17} can be rewritten in the 
dimensionless simple form
\begin{equation}
\tilde{\mathbf{q}} = \frac{1}{\mu} \Bigl[ - \frac{A_d}{2\gamma F}\, 
\Bigl\{ 1 + \frac{f}{2} A_1 \Bigr\} \, \nabla \Delta \rho 
- \mathrm{Im}[ Y^* \nabla Y] \Bigr] \ .
\label{Eq:C35}
\end{equation}
The overall factor $1/\mu$ is needed to keep the nabla operators dimensionless. We 
note that $A_d$ is a geometrical factor related to surface of the $d$ dimensional 
unit sphere \cite{Do85}. It should not be confused with the amplitudes $A$ and $A_1$. 
In an analogous way Eqs.\ \eqref{Eq:C06} and \eqref{Eq:C26} for the modified 
temperature parameter $\rho_1$ and its derivative $\nabla \rho_1$ can be written 
in a dimensionless form. We obtain
\begin{eqnarray}
\rho_1 \{ 1 + 8u A \} &=& \rho + (4u/A_d) \, |Y|^2 \ ,
\label{Eq:C36} \\
(\nabla\rho_1) \{ 1 + 8u A_1 \} &=& \nabla\rho + (4u/A_d) \, \nabla |Y|^2 \ , \qquad
\label{Eq:C37}
\end{eqnarray}
where the second equation should be multiplied by an overall factor $1/\mu$ 
to make the nabla operators dimensionless. Finally, the renormalized specific heat 
$C_\mathrm{ren}$ defined in \eqref{Eq:C22} and the parameter $X$ defined in 
\eqref{Eq:C18} are already dimensionless, so that we can keep them unchanged. 
We must only insert the dimensionless temperature parameters 
\eqref{Eq:C27}-\eqref{Eq:C29} and use the dimensionless nabla operator 
$\mu^{-1}\nabla$.

Now, all variables and parameters are expressed in a dimensionless form. Thus,
we are ready to rewrite the renormalized model-$F$ equations in dimensionless forms. 
From Eqs.\ \eqref{Eq:C19} and \eqref{Eq:C24} we obtain
\begin{eqnarray}
\frac{1}{g\mu^2} \, \frac{\partial Y}{\partial t} &=& 
- \frac{w}{F} \bigl[ \rho_1 - \mu^{-2}\nabla^2 \bigr] Y \nonumber\\ 
&&+ \frac{i}{2 \gamma} \Delta \rho \, Y \ , \qquad
\label{Eq:C38} \\
\frac{C_\mathrm{ren}}{2\gamma} \frac{A_d}{g\mu^2}
\, \frac{\partial \Delta \rho}{\partial t} 
&=& - \mu^{-1}\nabla \tilde{\mathbf{q}} \ ,
\label{Eq:C39}
\end{eqnarray}
where $w=w^\prime + i w^{\prime\prime}$ is a complex parameter.
In these equations we clearly see that $\mu$ defines the length scale 
and $g \mu^2$ defines the time scale. We may interpret $\mu^{-1}\nabla$ 
as a dimensionless nabla operator and $(g \mu^2)^{-1} \partial / \partial t$ 
as a dimensionless time derivative.

\subsection{Evaluation of the perturbative amplitudes}
\label{Sec:03C}
The amplitudes $A$ and $A_1$, defined in \eqref{Eq:C33} and \eqref{Eq:C34}, 
respectively, represent the contributions of the perturbation series expansion 
which in our case is the Hartree term. In order to solve the model-$F$ equations 
as partial differential equations we must have explicit expression to evaluate 
these amplitudes. The non-trivial contribution in the amplitudes is the function 
$\Phi_\alpha(X)$ together with its variable $X$ which are defined in 
Eqs.\ \eqref{Eq:B25} and \eqref{Eq:C18}. This function was first derived in 
Ref.\ \onlinecite{HD91}. Unfortunately, the function is a divergent infinite 
series so that it is not well defined in this form. However, in thermal equilibrium 
at zero heat current $Q=0$ and zero gravity $g=0$ this function can be omitted 
because it is just unity. In this case all temperature gradients are zero, so 
that the variable $X$ is zero which implies $\phi_\alpha(X=0)=1$. Hence, the 
amplitudes reduce to 
\begin{eqnarray}
A &=& \varepsilon^{-1}[ \rho_1^{-\varepsilon/2} - 1 ] \ ,
\label{Eq:C40} \\
A_1 &=& \varepsilon^{-1}[ (1 - \varepsilon/2) \rho_1^{-\varepsilon/2} - 1 ] \ .
\label{Eq:C41}
\end{eqnarray}
Since $\rho_1\sim \rho\sim (T-T_\lambda)$ must be positive, these amplitudes are 
valid for the \emph{normal-fluid} equilibrium state only and agree with former 
results \cite{Do85}.

In the nonequilibrium state the variable $X$ is nonzero. In this case the 
infinite series \eqref{Eq:B25} must be resummed to obtain a well-defined 
expression which can be evaluated numerically. Following our previous papers 
\cite{HD91,Ha99a} we write
\begin{equation}
\Phi_\alpha(X) = [\Gamma(\alpha)]^{-1} \, \zeta^\alpha \, 
\mathcal{F}_\alpha(\zeta)
\label{Eq:C42}
\end{equation}
where $\zeta=(-X)^{-1/3}$ and 
\begin{equation}
\mathcal{F}_\alpha(\zeta) = \int_0^\infty dv \, v^{\alpha-1} \, 
\exp(-v^3 - v\zeta) \  .
\label{Eq:C43}
\end{equation}
The integral is well defined for $\alpha>0$ and obtained by analytical continuation 
for $\alpha<0$. The new variable $\zeta$ is defined by a third root. Consequently, 
$\zeta$ is not unique a priori and may be complex. We must specify the root which 
should be taken. For this purpose we define the dimensionless parameter
\begin{eqnarray} 
\sigma &=& - \frac{1}{12 \, \mu^2} \Bigl[ (\nabla \rho_1)^2 
+ 2 \frac{w^{\prime\prime}}{w^\prime} 
\Bigl( \frac{F}{4 \gamma w^\prime} \nabla \Delta \rho \Bigr) 
\cdot \nabla \rho_1 \nonumber\\
&&- \Bigl( \frac{F}{4 \gamma w^\prime} \nabla \Delta \rho \Bigr)^2
\Bigr] \ .
\label{Eq:C44} 
\end{eqnarray}
so that $X=-\sigma/\rho_1^3$. Hence, the new variable can be written in the form 
$\zeta = \rho_1/\sigma^{1/3}$.

The transition from normal-fluid to superfluid $^4$He is related to a change of 
sign of $\rho_1\sim\rho\sim(T-T_\lambda)$. Consequently, also the new variable 
$\zeta$ changes sign. The nontrivial third root to be evaluated is $\sigma^{1/3}$. 
For this reason, we must distinguish two cases which are related to the two possible 
signs of $\sigma$. This distinction has important physical consequences. There will 
be two kinds of nonequilibrium superfluid phases of liquid $^4$He which are related 
to the two regimes where either the heat current $Q$ or the gravity $g$ is the 
dominating external influence. We discuss these two cases in the following.

\subsubsection{Heat current dominated regime: $\sigma>0$}
\label{Sec:03C1}
The self-organized critical state observed in the experiment by Moeur 
\emph{et al.}\ \cite{Du97} implies linear temperature profiles $T(z)$ and 
$T_\lambda(z)$ as function of the altitude $z$. The temperature difference 
$T(z)-T_\lambda(z)=\Delta T$ is constant over a large range of the altitude. 
Consequently, the related gradient parameters 
$\nabla \rho_1\sim \nabla \rho\sim \nabla (T-T_\lambda)$ are zero. On the other 
hand the heat current $\mathbf{Q}$ causes a nonzero constant gradient 
$\nabla \Delta\rho\sim \nabla T$. Thus, in the formula \eqref{Eq:C44} only the 
last term is nonzero which yields a positive result for $\sigma$. Hence, for 
the self-organized critical state the dimensionless parameter $\sigma$ is 
always constant and positive.

For the inhomogeneous nonequilibrium state we may conclude that $\sigma$ is also 
positive whenever the heat current $\mathbf{Q}$ and hence the related gradient 
$\nabla \Delta\rho$ is large compared to the effects of gravity. In our previous 
paper \cite{Ha99a} we confirm $\sigma>0$ for heat transport in liquid $^4$He on 
earth for heat currents $Q\gtrsim 70\,\mathrm{nW/cm^2}$. Moreover, for an 
experiment in zero gravity conditions in space, $\sigma$ is positive for all 
heat currents.

Whenever $\sigma$ is positive, the root $\sigma^{1/3}$ is straight forward. We 
just take the real positive root. Consequently, the variable $\zeta$ is real 
and changes sign at the superfluid transition. We find $\zeta>0$ in the normal 
fluid regime and $\zeta<0$ is the superfluid regime. The function \eqref{Eq:C42} 
and the integral \eqref{Eq:C43} can be evaluated directly. As a result we obtain 
the amplitudes
\begin{eqnarray}
A &=& \frac{1}{\varepsilon} \Bigl[ 
\frac{\sigma^{-\varepsilon/6}}{\Gamma(-1+\varepsilon/2)} 
\, \zeta^{-1} \mathcal{F}_{-1+\varepsilon/2}(\zeta) - 1 \Bigr] \ , \qquad
\label{Eq:C45} \\
A_1 &=& \frac{1}{\varepsilon} \Bigl[ 
- \frac{\sigma^{-\varepsilon/6}}{\Gamma(-1+\varepsilon/2)} 
\, \mathcal{F}_{\varepsilon/2}(\zeta) - 1 \Bigr] \ ,
\label{Eq:C46}
\end{eqnarray}
which we have derived and used in our previous paper \cite{Ha99a}.

We investigate the asymptotic behaviors of the function 
$\mathcal{F}_\alpha(\zeta)$ and find
\begin{equation}
\mathcal{F}_\alpha(\zeta) \approx \Gamma(\alpha) \, \zeta^{-\alpha}
\label{Eq:C47}
\end{equation}
for $\zeta\gg+1$ in the normal-fluid regime and
\begin{equation}
\mathcal{F}_\alpha(\zeta) \approx (\pi/3)^{1/2} \, (-\zeta/3)^{\alpha/2-3/4}
\, \exp\{ 2 (-\zeta/3)^{3/2} \}
\label{Eq:C48}
\end{equation}
for $\zeta\ll-1$ in the superfluid regime, respectively \cite{Ha99a}. In the first 
asymptotic case \eqref{Eq:C47} we recover the amplitudes \eqref{Eq:C40} and 
\eqref{Eq:C41} of the normal-fluid equilibrium state. In the second asymptotic case 
\eqref{Eq:C48} we obtain exponentially large amplitudes $A$ and $A_1$ for the 
nonequilibrium superfluid state.

The latter asymptotic case has an important physical consequence. We consider 
Eq.\ \eqref{Eq:C36} which is a constraint to define $\rho_1$. In the original 
form related to \eqref{Eq:B16} this equation is rewritten as
\begin{equation}
\rho_1 = \rho - 8u A \rho_1 + (4u/A_d) \, |Y|^2 \ .
\label{Eq:C49}
\end{equation}
The last term is the contribution of the renormalized complex order parameter $Y$ 
which is nonzero only in the superfluid state. However, in the superfluid regime 
the second term may be a competing term because the amplitude $A$ may be 
exponentially large. Thus, in the nonequilibrium system there may be two 
competing superfluid phases which have different physical properties. In Eq.\ 
\eqref{Eq:B22} we have split the order-parameter Green function into two terms,
a mean-field term and a fluctuating term. The third and the second term in 
\eqref{Eq:C49} refer to these two terms of the Green function, respectively.

The complex order parameter $Y$ may be decomposed into a modulus $\eta$ and 
a phase $\varphi$ according to $Y=\eta\, e^{i\varphi}$. In our previous papers 
\cite{Ha99,Ha99a} we argue that in the superfluid regime the modulus $\eta$ 
and hence the average order parameter $Y$ is zero because of large fluctuations of 
the phase $\varphi$. These large phase fluctuations are related to vortices 
and quantum turbulence. In the present paper we consider a nonzero average order 
parameter $Y$ in the superfluid regime and solve the renormalized model-$F$ equations 
numerically as partial differential equations. We find a competition between the 
\emph{mean-field superfluid phase}, described by the average order parameter $Y$, 
and the \emph{fluctuating superfluid phase}, described by the exponentially large 
amplitude $A$.

\subsubsection{Gravity dominated regime: $\sigma<0$}
\label{Sec:03C2}
In thermal equilibrium for zero heat currents $Q=0$ the temperature $T$ is
constant. Consequently, the gradient $\nabla \Delta \rho\sim \nabla T$ is zero.
On the other hand, gravity on earth implies a nonzero gradient of the 
critical temperature $\nabla T_\lambda$. Hence the other gradients 
$\nabla \rho_1\sim \nabla \rho \sim \nabla (T-T_\lambda)=-\nabla T_\lambda$ 
are nonzero. In Eq.\ \eqref{Eq:C44} only the first term is nonzero which implies 
a negative dimensionless parameter $\sigma$. A small heat current $Q$ will not 
change the situation. In our numerical calculations we find $\sigma<0$ for 
$Q\lesssim 20\,\mathrm{nW/cm^2}$. 

An exception is the self-organized critical state which always implies $\sigma > 0$
and which exists for arbitrary small heat currents $Q$ where the temperature difference 
$T(z)-T_\lambda(z)=\Delta T$ is constant. Nevertheless, for small heat currents 
$\Delta T$ is positive so that the system is normal fluid and the sign of $\sigma$ 
is irrelevant.

For negative $\sigma$ the third root is always complex. We find 
$\sigma^{1/3}=e^{\mp i\pi/3} (-\sigma)^{1/3}$, so that the variable of the function 
\eqref{Eq:C43} is complex, i.e.\ $\zeta = e^{\pm i\pi/3} \rho_1 / (-\sigma)^{1/3}$.
For convenience we introduce the new real parameter 
$\bar\zeta = \rho_1 / (-\sigma)^{1/3}$ which is related to the old parameter via 
$\zeta = e^{\pm i\pi/3} \bar\zeta$. We furthermore define the new complex function
\begin{equation}
\mathcal{G}_\alpha(\bar\zeta) = e^{\pm i\alpha\pi/3} \, \mathcal{F}_\alpha(\zeta)
= e^{\pm i\alpha\pi/3} \, \mathcal{F}_\alpha(e^{\pm i\pi/3} \bar\zeta) \ .
\label{Eq:C50}
\end{equation}
which can be decomposed into real and imaginary parts according to 
$\mathcal{G}_\alpha(\bar\zeta) = \mathcal{G}^\prime_\alpha(\bar\zeta) 
\pm i \, \mathcal{G}^{\prime\prime}_\alpha(\bar\zeta)$. The new complex function 
is not uniquely defined because there are two complex roots which can be chosen. 
This fact causes two possible signs for the imaginary part. However, we choose 
the so called \emph{principal} part, which is obtained as the average of the two 
cases so that the imaginary part cancels. Thus, we simply omit the imaginary part 
$\mathcal{G}^{\prime\prime}_\alpha(\bar\zeta)$. In the normal-fluid region 
$\bar\zeta>0$ this assumption is plausible because the imaginary part converges
to zero exponentially for increasing $\bar\zeta$. As a result, we rewrite the 
amplitudes \eqref{Eq:C45} and \eqref{Eq:C46} in terms of the new function 
\eqref{Eq:C50} as
\begin{eqnarray}
A &=& \frac{1}{\varepsilon} \Bigl[ 
\frac{(-\sigma)^{-\varepsilon/6}}{\Gamma(-1+\varepsilon/2)} 
\, \bar\zeta^{-1} \mathcal{G}^\prime_{-1+\varepsilon/2}(\bar\zeta) - 1 \Bigr] \ , \qquad
\label{Eq:C51} \\
A_1 &=& \frac{1}{\varepsilon} \Bigl[ 
- \frac{(-\sigma)^{-\varepsilon/6}}{\Gamma(-1+\varepsilon/2)} 
\, \mathcal{G}^\prime_{\varepsilon/2}(\bar\zeta) - 1 \Bigr] \ .
\label{Eq:C52}
\end{eqnarray}
Once again, we consider the asymptotic behaviors of the function 
$\mathcal{G}^\prime_\alpha(\bar\zeta)$. We find
\begin{equation}
\mathcal{G}^\prime_\alpha(\bar\zeta) \approx \Gamma(\alpha) 
\, \bar\zeta^{-\alpha}
\label{Eq:C53}
\end{equation}
for $\bar\zeta\gg+1$ in the normal-fluid regime and
\begin{eqnarray}
\mathcal{G}^\prime_\alpha(\bar\zeta) &\approx& (\pi/3)^{1/2} 
\, (-\bar\zeta/3)^{\alpha/2-3/4} \nonumber\\
&&\times \cos\{ 2 (-\bar\zeta/3)^{3/2} + (\pi/4)(2\alpha-1) \} \qquad
\label{Eq:C54}
\end{eqnarray}
for $\bar\zeta\ll-1$ in the superfluid regime. In the first asymptotic case 
\eqref{Eq:C53} we recover the amplitudes \eqref{Eq:C40} and \eqref{Eq:C41} of 
the normal-fluid equilibrium state. Since here the amplitudes do not depend 
on the dimensionless parameter $\sigma$ at all, in the normal-fluid regime 
the sign of $\sigma$ is irrelevant. In the second asymptotic case \eqref{Eq:C54}
the function $\mathcal{G}_\alpha(\bar\zeta)$ and hence the amplitudes $A$ and 
$A_1$ oscillate but remain of order unity.

Again, the latter asymptotic case has an important physical consequence. 
In Eq.\ \eqref{Eq:C49} the second term is always small because the amplitude 
$A$ never becomes large. Hence the superfluid phase is unique. It is the 
\emph{mean-field superfluid phase} where the average order parameter $Y$ is 
nonzero. Vortices due to fluctuations effects and a fluctuating superfluid 
phase do not exist for $\sigma<0$.

\subsection{Renormalization-group theory \newline and flow parameter condition}
\label{Sec:03D}
In the renormalization procedure the parameter $\mu$ is introduced which fixes 
the length scale. This parameter generates a transformation group which is known 
as the renormalization group. Following Ref.\ \onlinecite{Do85} it can be changed 
by the substitution $\mu\to\mu\ell$, where the dimensionless parameter $\ell$ is 
called the renormalization-group (RG) flow parameter \cite{Do85}. However, for 
simplicity and consistency of the following calculations, in this paper we do not 
use the above substitution. We avoid the use of the flow parameter $\ell$ and 
thus change the length scale parameter $\mu$ directly. We use the alternative 
dimensionless RG flow parameter $\tau$ which is defined in \eqref{Eq:C30}.
All quantities of the renormalized theory can be expressed in terms of this RG 
flow parameter. The dimensionless coupling parameters are $u[\tau]$, 
$\gamma[\tau]$, $w[\tau]=w^\prime[\tau]+iw^{\prime\prime}[\tau]$, $F[\tau]$, 
and $f[\tau]$. This is a notation which was defined in Refs.\ \onlinecite{Do85} 
and \onlinecite{Do91}. 

A differential relation between the flow parameter $\tau$ and the length-scale 
parameter $\mu$ can be obtained by a logarithmic differentiation of 
Eq.\ \eqref{Eq:C30}, which reads
\begin{equation}
d\ln\tau = \Bigl[ \frac{1}{2} \Bigl( d + \frac{\partial\ln Z_m^{-1}}{\partial\ln\mu} \Bigr) 
- \frac{\partial\ln\gamma}{\partial\ln\mu} \Bigr]\, d\ln\mu \ .
\label{Eq:C55}
\end{equation}
Using the definitions of the RG zeta functions \cite{Do85}
\begin{eqnarray}
\zeta_\phi &=& \partial\ln Z_\phi^{-1} / \partial \ln\mu \ , 
\label{Eq:C56} \\
\zeta_r &=& \partial\ln Z_r^{-1} / \partial \ln\mu \ , 
\label{Eq:C57} \\
\zeta_m &=& \partial\ln Z_m^{-1} / \partial \ln\mu \ , 
\label{Eq:C58}
\end{eqnarray}
using the RG equation for the parameter $\gamma$ \cite{Do85}
\begin{equation}
\partial\ln\gamma / \partial\ln\mu = [ - \varepsilon + 2\zeta_r + \zeta_m ] / 2 \ ,
\label{Eq:C59}
\end{equation}
and using $\varepsilon=4-d$, Eq.\ \eqref{Eq:C55} can be simplified into
\begin{equation}
d\ln\tau = [ 2 - \zeta_r ] \, d\ln\mu \ .
\label{Eq:C60}
\end{equation}
The zeta function $\zeta_r=\zeta_r(u)$ is explicitly available as a function of 
$u=u[\tau]$ \cite{SD89}. Thus Eq.\ \eqref{Eq:C60} enables an explicit numerical 
calculation of $\tau$ as a function of $\mu$ and vice versa.

Since the renormalization procedure implies a reordering of the perturbation 
series, the RG flow parameter $\tau$ should be chosen in an optimum way so that 
the convergence behavior of the series is optimized. To do this we choose the 
constraint condition 
\begin{equation}
3\,\rho_1 - 2\,\rho + 3 (4u/A_d) f_Y ( \nabla Y / \mu )^2 
+ f_{\Delta\rho} ( \nabla \Delta\rho / \mu)^2 = 1 \ . 
\label{Eq:C61}
\end{equation}
The modified temperature parameter $\rho_1$ is defined in Eq.\ \eqref{Eq:C36} 
which may be viewed as a second constraint equation. The first two terms on the 
left-hand side of \eqref{Eq:C61} guarantee the standard flow parameter conditions 
of normal-fluid and superfluid $^4$He in thermal equilibrium and zero gravity 
which have been formulated in Ref.\ \onlinecite{Do85}. The latter two terms 
are gradient terms which stabilize the intermediate region of the 
superfluid/normal-fluid interface. The two parameters $f_Y$ and $f_{\Delta\rho}$ 
are dimensionless and control the influence of the gradient terms. In our 
calculations we have used $f_Y=5$ and $f_{\Delta\rho}=1$ as an optimum choice.

In thermal equilibrium and zero gravity all quantities and parameters are 
constant in space and time. An exception is the renormalized order parameter 
$Y=Y(t)=\eta\, e^{i\varphi(t)}$ together with the constant modulus $\eta$ and 
the time-dependent phase $\varphi(t) = - \omega t + \varphi_0$. Since all gradient 
terms are zero, the model-$F$ equations \eqref{Eq:C38}, \eqref{Eq:C39} and the 
flow-parameter equation \eqref{Eq:C61} reduce to 
\begin{eqnarray}
\omega &=& - g\mu^2 (2\gamma)^{-1} \, \Delta\rho \ ,
\label{Eq:C62} \\
\rho_1\, Y &=& 0 \ ,
\label{Eq:C63} \\
3\,\rho_1 - 2\,\rho &=& 1 \ .
\label{Eq:C64}
\end{eqnarray}
The first equation is always satisfied because it defines the order-parameter 
frequency $\omega$ in terms of the dimensionless renormalized temperature 
difference $\Delta\rho$ where the time scale is ruled by the parameter 
combination $g\mu^2$.

In the normal-fluid state, the second equation \eqref{Eq:C63} implies the 
zero order parameter $Y=0$, where $\rho_1$ may be nonzero. The flow-parameter 
equation \eqref{Eq:C64} together with the constraint \eqref{Eq:C36} and 
the amplitudes \eqref{Eq:C40} and \eqref{Eq:C41} imply $\rho=\rho_1=1$, 
$A=0$, and $A_1=-1/2$. These results are compatible with the equilibrium 
theory of Ref.\ \onlinecite{Do85}. The resulting flow parameter condition is
$\rho=1$ which in the notation of Ref.\ \onlinecite{Do85} reads 
$r(l)/(\mu\ell)^2=1$. Consequently, from Eq.\ \eqref{Eq:C28} we obtain 
the flow parameter $\tau=(T-T_\lambda)/T_\lambda$ which just is the reduced 
temperature as known from earlier work \cite{Do85}.

In the superfluid state, Eq.\ \eqref{Eq:C63} implies $\rho_1=0$ where the 
order parameter $Y$ is nonzero. Consequently, Eq.\ \eqref{Eq:C64} yields 
the flow-parameter condition $\rho=-1/2$ which is well known from 
Ref.\ \onlinecite{Do85} in the notation $r(\ell)/(\mu\ell)^2=-1/2$. Again, 
Eq.\ \eqref{Eq:C28} relates the flow parameter to the reduced temperature according 
to $\tau=2(T_\lambda-T)/T_\lambda$. From the constraint \eqref{Eq:C36} we obtain 
the modulus of the order parameter $\eta=|Y|$. Since the left-hand side is zero, 
we obtain $\eta=(A_d/8u)^{1/2}$.

The above investigation of the normal-fluid and superfluid equilibrium states in 
zero gravity shows, that in our numerical calculations for the superfluid/normal-fluid 
interface the dimensionless renormalized temperature variables $\rho$, $\rho_1$, 
and the modulus of the dimensionless renormalized order parameter $\eta=|Y|$ must 
approach constant asymptotic values on both sides far away from the interface. In 
the intermediate region near the interface, the variables will interpolate the 
asymptotic values. The RG flow-parameter condition \eqref{Eq:C61} guarantees the 
asymptotic values and yields an appropriate interpolation in the intermediate 
interface region. The gradient terms in this condition will stabilize the interpolation.

The RG flow-parameter condition \eqref{Eq:C61} is designed for the 
superfluid/normal-fluid interface at small heat currents where gravity is 
the dominating external influence and where in the superfluid phase the order
parameter $Y$ is nonzero. In our classification of Sec.\ \ref{Sec:03C} this
superfluid phase is the mean-field superfluid phase. The other case is the
fluctuating superfluid phase where the order parameter $Y$ is zero and vortices 
are present. This latter case has been investigated in our previous publications 
\cite{Ha99,Ha99a} where the RG flow-parameter condition is given by Eqs.\ (11) 
and (4.39) of Refs.\ \onlinecite{Ha99} and \onlinecite{Ha99a}, respectively. 
This latter flow parameter condition can be compared with our present condition 
\eqref{Eq:C61} if $\rho$ is eliminated by using the second constraint \eqref{Eq:C36} 
and if we use $\rho_1=\sigma^{1/2}\zeta$. Then, the first and second term 
of our present condition \eqref{Eq:C61} are identified with the second and third 
term in Eqs.\ (11) and (4.39) of Ref.\ \onlinecite{Ha99} and \onlinecite{Ha99a}. 
The gradient terms of Eq.\ \eqref{Eq:C61} are replaced by the first term in 
Eqs.\ (11) and (4.39) of Refs.\ \onlinecite{Ha99} and \onlinecite{Ha99a}, which 
is also a gradient term because $\sigma$ is depends on the gradients following 
\eqref{Eq:C44}. We note that the RG flow-parameter condition of 
Ref.\ \onlinecite{Ha99} and \onlinecite{Ha99a} is designed for the fluctuating 
superfluid phase where the order parameter $Y$ is zero and vortices are present.

\subsection{Covariant derivatives}
\label{Sec:03E}
The flow parameter equation \eqref{Eq:C61} and the constraint condition \eqref{Eq:C36} 
are local equations. Consequently, the flow parameter $\tau$, the renormalization $Z$ 
factors, and the dimensionless coupling parameters are local and depend on space 
and time. This fact will affect the space and time derivatives in the renormalized 
equations. We must replace the partial differential operators by covariant 
derivatives. To do this, we write the renormalization equations in a form like 
Eqs.\ \eqref{Eq:C01}-\eqref{Eq:C03}, so that the bare quantities are on the left-hand 
side and all renormalized quantities are on the right hand side. Then we apply the 
differential operator. We start with the renormalization of the temperature 
\eqref{Eq:C27} which is equivalent to \eqref{Eq:C02}. We apply the nabla operator 
and obtain
\begin{eqnarray}
\nabla [(T-T_0)/T_\lambda] &=& \nabla [\tau\, \Delta\rho] 
= \tau \, [ \nabla + (\nabla\ln\tau) ] \Delta\rho \nonumber\\
&=& \tau \, \mathbf{D} \Delta\rho \ .
\label{Eq:C65}
\end{eqnarray}
The last equality sign defines the covariant derivative. We continue with the 
temperature difference \eqref{Eq:C28} which is equivalent to \eqref{Eq:C03} 
and proceed in the same way. As a result we obtain the covariant derivatives 
\begin{eqnarray}
\mathbf{D} \Delta\rho &=& [ \nabla + (\nabla\ln\tau) ] \Delta\rho \ ,
\label{Eq:C66} \\
\mathbf{D} \rho &=& [ \nabla + (\nabla\ln\tau) ] \rho \ .
\label{Eq:C67}
\end{eqnarray}
The covariant derivative of $\rho_1$ is more complicated. Generalizing 
Eq.\ \eqref{Eq:C37} we obtain
\begin{equation}
(\mathbf{D}\rho_1) \{ 1 + 8u A_1 \} = \mathbf{D}\rho 
+ (4u/A_d) \, \mathbf{D} |Y|^2
\label{Eq:C68}
\end{equation}
which can be resolved with respect to $\mathbf{D}\rho_1$.

Next we consider the renormalization of the order parameter \eqref{Eq:C01}.
We write this equation in terms of the dimensionless renormalized order parameter 
$Y$ by using \eqref{Eq:C31}, apply the nabla operator, use \eqref{Eq:C56}, and obtain
\begin{eqnarray}
\nabla \langle\psi\rangle &=& \nabla [ Z_\phi^{1/2} \mu^{(d-2)/2}\, Y ]
\nonumber\\
&=& Z_\phi^{1/2} \mu^{(d-2)/2}\, \Bigl[ \nabla + \frac{1}{2}
( d - 2 - \zeta_\phi ) (\nabla\ln\mu) \Bigr] \, Y \nonumber\\
&=& Z_\phi^{1/2} \mu^{(d-2)/2}\, \Bigl[ \nabla + \frac{1}{2}
\frac{ d - 2 - \zeta_\phi }{ 2 - \zeta_r } 
(\nabla\ln\tau) \Bigr] \, Y \nonumber\\
&=& Z_\phi^{1/2} \mu^{(d-2)/2}\, \mathbf{D} Y \ .
\label{Eq:C69}
\end{eqnarray}
Again, the last equality sign defines the covariant derivative. We define 
the running critical exponents \cite{SD89}
\begin{eqnarray}
\nu &=& 1 / ( 2 - \zeta_r ) \ ,
\label{Eq:C70} \\
\eta &=& - \zeta_\phi \ ,
\label{Eq:C71} \\
\beta &=& \nu ( d - 2 + \eta ) / 2 \ .
\label{Eq:C72}
\end{eqnarray}
These exponents are called \emph{running} exponents because they depend on the 
RG flow parameter $\tau$ via the zeta functions and thus carry all the Wegner 
corrections. In the asymptotic limit $\tau\to 0$ they converge to the universal 
critical exponents. Then from Eq.\ \eqref{Eq:C69} we obtain the covariant derivative 
of the dimensionless renormalized order parameter
\begin{equation}
\mathbf{D} Y = [ \nabla + \beta (\nabla\ln\tau) ] Y \ .
\label{Eq:C73}
\end{equation}
We note that $\beta$ is the running critical exponent of the order parameter.
This result makes clear, how the general structure of a covariant derivative of 
a dimensionless renormalized quantity looks like: It is the partial derivative 
of the quantity plus the critical exponent times the partial derivative of $\ln\tau$ 
times the quantity. In the model-$F$ equations we also need the second covariant 
derivative of the order parameter. It is obtained by applying the operator twice, 
i.e.\ 
\begin{equation}
\mathbf{D}^2 Y = [ \nabla + \beta (\nabla\ln\tau) ]^2 Y \ .
\label{Eq:C74}
\end{equation}

Furthermore we consider the renormalization of the heat current \eqref{Eq:C32}. 
Applying the nabla operator we obtain
\begin{eqnarray}
\nabla [ \mathbf{Q}/g_0 k_B T_\lambda ] &=& \nabla[ \mu^{d-1} \, \tilde{\mathbf{q}} ] 
\nonumber\\
&=& \mu^{d-1} \, [ \nabla + (d-1) (\nabla\ln\mu) ] \, \tilde{\mathbf{q}}
\nonumber\\
&=& \mu^{d-1} \, [ \nabla + (d-1) \nu (\nabla\ln\tau) ] \, \tilde{\mathbf{q}}
\nonumber\\
&=& \mu^{d-1} \, \mathbf{D} \tilde{\mathbf{q}} \ .
\label{Eq:C75}
\end{eqnarray}
Thus, we find the covariant derivative of the dimensionless renormalized heat current
\begin{equation}
\mathbf{D} \tilde{\mathbf{q}} = [ \nabla + (d-1)\nu (\nabla\ln\tau) ] 
\tilde{\mathbf{q}} \ .
\label{Eq:C76}
\end{equation}
We identify $(d-1)\nu$ as the running critical exponent of the heat current. The 
inverse exponent $x=1/[(d-1)\nu]$ is known from the depression of the critical 
temperature $T_\lambda$ by a nonzero heat current $Q$ \cite{On83,HD92}.
We note that the covariant derivatives \eqref{Eq:C66}-\eqref{Eq:C68} and 
\eqref{Eq:C73}-\eqref{Eq:C74} have been derived already in our previous unpublished 
approach \cite{HN08} for the interface in thermal equilibrium at zero heat current.

Above, we have defined the covariant derivatives with respect to the space coordinates 
$\mathbf{D}$. We also need the covariant derivatives with respect to time $D_t$. 
To obtain them we replace the nabla operator by the partial time derivative 
$\partial_t=\partial/\partial t$. Thus, as results we obtain e.g.\ 
\begin{eqnarray}
D_t Y &=& [ \partial_t + \beta (\partial_t \ln\tau) ] Y \ ,
\label{Eq:C77} \\
D_t \Delta\rho &=& [ \partial_t + (\partial_t \ln\tau) ] \Delta\rho \ .
\label{Eq:C78}
\end{eqnarray}

Now, we are ready to rewrite the model-$F$ equations in terms of covariant 
derivatives. From Eqs.\ \eqref{Eq:C38} and \eqref{Eq:C39} we obtain
\begin{eqnarray}
\frac{1}{2\gamma\, \tau} \, \frac{1}{g_0} \, D_t Y 
&=& - \frac{w}{F} \bigl[ \rho_1 - \xi^2 \mathbf{D}^2 \bigr] Y \nonumber\\
&&+ \frac{i}{2 \gamma} \Delta \rho \, Y \ ,
\label{Eq:C79} \\
\frac{C_\mathrm{ren}}{(2\gamma)^2\tau} \frac{A_d}{g_0}
\, D_t \Delta \rho &=& - \xi\, \mathbf{D} \tilde{\mathbf{q}} \ .
\label{Eq:C80}
\end{eqnarray}
In these equations we have performed some further substitutions which 
are known from our previous paper \cite{Ha99a}, i.e.\ 
\begin{eqnarray}
\mu = \xi^{-1} \ , \qquad  g \mu^2 = g_0 \, 2 \gamma \, \tau \ .
\label{Eq:C81}
\end{eqnarray}
Here $\xi=\xi[\tau]$ is the correlation length. Close to criticality it has the 
asymptotic form $\xi=\xi_0\, \tau^{-\nu}$. The identification $\mu=\xi^{-1}$ is 
correct in our Hartree approximation which is a self-consistent one-loop 
approximation. Corrections appear in higher orders \cite{SD89}. The renormalized 
time-scale parameter $g \mu^2$ is expressed in terms of the bare parameter $g_0$ 
by using the renormalization equation \eqref{Eq:C10} where $\chi_0 Z_m$ has been 
eliminated in favor of $\tau$ by \eqref{Eq:C30}. As a result the renormalized 
model-$F$ equations \eqref{Eq:C79} and \eqref{Eq:C80} are dimensionless equations 
for the dimensionless quantities. There are two parameters, which control the 
scales of space and time. They are $\xi_0$ and $g_0$, respectively.

The model-$F$ equations \eqref{Eq:C79} and \eqref{Eq:C80} are supplemented by some 
further equations including the dimensionless renormalized entropy current 
\eqref{Eq:C35} and the two constraint conditions \eqref{Eq:C36} and \eqref{Eq:C61} 
where all nabla operators $\nabla$ are replaced by respective covariant derivatives 
$\mathbf{D}$. Thus, we obtain the dimensionless renormalized entropy current
\begin{equation}
\tilde{\mathbf{q}} = - \frac{A_d}{2\gamma F}\, \Bigl\{ 1 + \frac{f}{2} A_1 \Bigr\} 
\, \xi\, \mathbf{D} \Delta \rho - \mathrm{Im}[ Y^* \xi\, \mathbf{D} Y] \ ,
\label{Eq:C82}
\end{equation}
the constraints
\begin{eqnarray}
K_1 &=& 3\,\rho_1 - 2\,\rho + 3 (4u/A_d) f_Y ( \xi\,\mathbf{D} Y )^2 \nonumber\\
&&+ f_{\Delta\rho} ( \xi\,\mathbf{D} \Delta\rho )^2 - 1 = 0 \ ,
\label{Eq:C83} \\
K_2 &=& \rho_1 \{ 1 + 8u A \} - [ \rho + (4u/A_d) \, |Y|^2 ] = 0 \ , \qquad
\label{Eq:C84}
\end{eqnarray}
and furthermore the dimensionless variables
\begin{eqnarray} 
\sigma &=& - \frac{1}{12} \Bigl[ (\xi\,\mathbf{D} \rho_1)^2 
+ 2 \frac{w^{\prime\prime}}{w^\prime} 
\Bigl( \frac{F}{4 \gamma w^\prime} \xi\,\mathbf{D} \Delta \rho \Bigr) 
\cdot ( \xi\,\mathbf{D} \rho_1 ) \nonumber\\
&&- \Bigl( \frac{F}{4 \gamma w^\prime} \xi\,\mathbf{D} \Delta \rho \Bigr)^2
\Bigr] \ ,
\label{Eq:C85} 
\end{eqnarray}
and $\zeta = \rho_1/\sigma^{1/3}$ or $\bar\zeta=\rho_1 / (-\sigma)^{1/3}$ which 
are needed to calculate the dimensionless amplitudes \eqref{Eq:C45}-\eqref{Eq:C46} 
or \eqref{Eq:C51}-\eqref{Eq:C52}.

\subsection{Numerical algorithm}
\label{Sec:03F}
The numerical algorithm for solving the renormalized model-$F$ equations 
\eqref{Eq:C79} and \eqref{Eq:C80} together with the constraints \eqref{Eq:C83} 
and \eqref{Eq:C84} is implemented by two iterations. First on the left-hand sides 
of the model-$F$ equations the partial time derivatives within the covariant 
derivatives are replaced by discrete forward differences
\begin{eqnarray}
\partial_t Y &\to& [ Y(\mathbf{r},t+\Delta t) - Y(\mathbf{r},t) ] / \Delta t \ ,
\label{Eq:C86} \\
\partial_t \Delta\rho &\to& [ \Delta\rho(\mathbf{r},t+\Delta t) 
- \Delta\rho(\mathbf{r},t) ] / \Delta t \ .
\label{Eq:C87}
\end{eqnarray}
Secondly, the constraints are solved by a Newton method. The two iterations 
are performed in parallel, i.e.\ alternatively one time step and one Newton 
step. In this way starting with appropriate initial functions at an initial 
time the dimensionless renormalized quantities $Y(\mathbf{r},t)$, 
$\Delta\rho(\mathbf{r},t)$ and $\rho_1(\mathbf{r},t)$, $\ln\tau(\mathbf{r},t)$ 
are obtained as functions of space and time. All the other dimensionless renormalized 
quantities which are needed on the right and sides of the iteration equations 
can be calculated from the four quantities by formulas we have derived above. 
The covariant derivatives $\mathbf{D}Y$, $\mathbf{D}^2Y$, $\mathbf{D}\Delta\rho$,
$\mathbf{D}\rho$, and $\mathbf{D}\tilde{q}$ are calculated with discrete 
nabla and Laplace operators on an equidistant grid of the space-coordinates 
$\mathbf{r}$. The covariant derivatives of further quantities can by related to 
those five by equations like \eqref{Eq:C68}.

For the Newton-iteration step we need the derivatives of the constraint 
functions $K_1$ and $K_2$ with respect to $\rho_1$ and $\ln\tau$. We use 
the derivatives 
\begin{eqnarray}
\partial K_1 / \partial \rho_1 &=& 3 \ ,
\label{Eq:C88} \\
\partial K_2 / \partial \rho_1 &=& 1 + 8u\,A_1 \ ,
\label{Eq:C89} \\
\partial K_1 / \partial \ln\tau &\approx& - \bigl\{ -2\rho 
+ 3 (4u/A_d) f_Y ( \xi\, \mathbf{D} Y ) ^2 2 ( \beta + \nu ) \nonumber\\
&&+ f_{\Delta\rho} ( \xi\, \mathbf{D} \Delta\rho )^2 2 ( 1 + \nu ) \bigr\} \ ,
\label{Eq:C90} \\
\partial K_2 / \partial \ln\tau &\approx& - \bigl\{ 8u\, E_1 \, 2 ( 1 + \nu )
\nonumber\\
&&- [ \rho + (4u/A_d) Y^2 2\beta ] \bigr\}
\label{Eq:C91}
\end{eqnarray}
together with the amplitude
\begin{equation}
E_1 = (\rho_1 / 6) \bigl[ (2-\varepsilon) A - 2 A_1 - 1 \bigr] \ .
\label{Eq:C92}
\end{equation}
The latter two derivatives are approximations, because we omitted the weak 
dependence of the dimensionless coupling parameters $u[\tau]$, $\gamma[\tau]$, 
etc.\ on the logarithmic RG flow parameter $\ln\tau$. Nevertheless, our numerical 
calculation works. There is no significant influence of this approximation.

Our numerical calculations are performed very close to criticality where $\tau< 10^{-5}$. 
Consequently, for the running exponents we can use the universal critical 
exponents as a good approximation. We use the experimental value $\nu=0.671$ of 
Lipa \emph{et al.}\ \cite{Li96,Li03} and the theoretical value $\eta=0.038$ of Schloms 
and Dohm \cite{SD89}. The exponent $\beta=0.348$ is calculated from the scaling 
relation \eqref{Eq:C72}, where the dimension of space is $d=3$. Finally, we can 
use the asymptotic formula for the correlation length $\xi=\xi_0\, \tau^{-\nu}$ 
as a good approximation.

Our numerical calculations show that the iterations are stable for small heat 
currents $Q=|\mathbf{Q}|\lesssim 20\,\mathrm{nW/cm^2}$ where the gravity is the 
dominating external force and the dimensionless parameter $\sigma$ defined in 
\eqref{Eq:C85} is always negative. For lager heat currents the parameter $\sigma$ 
will have a sign change locally in space, which causes numerical troubles. We can 
stabilize the calculations up to a maximum heat current 
$Q_\mathrm{max}=160\,\mathrm{nW/cm^2}$ by adding a small imaginary constant 
to the right-hand side of Eq.\ \eqref{Eq:C85}. However, for larger heat currents 
where the heat flow is the major and the gravity is the minor external influence 
the iteration is unstable so that no results can be obtained.

\section{Numerical results}
\label{Sec:04}
Most experiments with liquid $^4$He close to the superfluid transition are performed 
at saturated vapor pressure. The temperature $T$ is varied in the region near 
$T_\lambda$ where the pressure is kept at the value of the liquid-gas transition. 
In this case the critical temperature is $T_\lambda=2.172\,\mathrm{K}$. The parameters 
which specify the scales of length and time are $\xi_0=1.44\times10^{-8}\,\mathrm{cm}$ 
and $g_0=2.164\times 10^{11}\,\mathrm{s^{-1}}$, respectively \cite{Al85,Li96,Li03}. The 
dimensionless renormalized coupling parameters $u[\tau]$, $\gamma[\tau]$, 
$w[\tau]=w^\prime[\tau]+i w^{\prime\prime}[\tau]$, $F[\tau]$, and $f[\tau]$ as functions 
of the RG flow parameter $\tau$ are taken from Ref.\ \onlinecite{Do91}.

We perform the numerical calculations for liquid $^4$He in $d=3$ dimensions. The 
system is assumed to be homogeneous in the two horizontal directions $x$ and $y$. 
Thus, all quantities and functions depend only on the altitude coordinate $z$ and 
the time $t$. The model-$F$ equations reduce to partial differential equations with 
the two variables $z$ and $t$. The size of the experimental cells which contain the 
liquid $^4$He is usually some millimeters in $z$ direction. We use a cell length 
$L=2.0\,\mathrm{mm}$ and discretize the $z$ coordinate into $500$ points. Consequently, 
the discretization is $\Delta z=4.0\,\mathrm{\mu m}$ in the altitude coordinate.

The discretization of the time $\Delta t$ in the partial derivatives \eqref{Eq:C86} 
and \eqref{Eq:C87} must be sufficiently small so that the iteration converges. On the 
other hand $\Delta t$ should be sufficiently large, so that the calculation time on 
the computer is not too long. We find $\Delta t = 4.0\times10^{-6}\,\mathrm{s}$ as 
an optimum choice. Starting the calculations in any nonequilibrium state, we first 
observe space and time dependent oscillations which are related to second sound. 
These oscillations relax on a time scale of about one second. After a time interval
$\delta t=2.0\,\mathrm{s}$ the system reaches a stationary state with a constant 
homogeneous heat current $\mathbf{Q}$ where all oscillations are disappeared. This 
means we need $5\times 10^5$ iteration steps on the computer until the system 
converges to the steady state.

For a heat flow in $z$ direction there must be a heat source and a heat sink 
at the boundaries of the cell $z_1=-L/2=-1.0\,\mathrm{mm}$ and 
$z_2=+L/2=+1.0\,\mathrm{mm}$, respectively. Thus, a source and sink term must 
be added to the heat transport equation of model $F$ \eqref{Eq:B02} which is
given by
\begin{equation}
W(\mathbf{r},t) = 2\, [ Q_1\,\delta(z-z_1) - Q_2\,\delta(z-z_2) ] \ .
\label{Eq:D01}
\end{equation}
In dimensionless renormalized form the source and sink term is
\begin{equation}
\tilde{w}(\mathbf{r},t) = 2\, [ \tilde{q}_1\,\delta([z-z_1]/\xi) 
- \tilde{q}_2\,\delta([z-z_2]/\xi) ] \ .
\label{Eq:D02}
\end{equation}
This latter term must be added to the dimensionless renormalized model-$F$ 
equation \eqref{Eq:C80} on the right-hand side. The relation between the 
dimensionless renormalized heat currents $\tilde{q}$ and the physical heat 
currents $Q$ is obtained from the renormalization equations \eqref{Eq:C08},
\eqref{Eq:C10}, and \eqref{Eq:C32}. We obtain
\begin{equation}
\tilde{q} = \frac{q}{g_0}\, \xi^{d-1}
= \frac{Q\, \xi^{d-1}}{g_0 k_\mathrm{B} T_\lambda}
\label{Eq:D03}
\end{equation}
which should be applied to both heat currents in \eqref{Eq:D01} and 
\eqref{Eq:D02}. It is important to note, that the RG flow parameter 
$\tau=\tau(\mathbf{r},t)$ and hence the correlation length $\xi=\xi[\tau]$ 
depend on space and time. This fact is important for Eqs.\ \eqref{Eq:D01} 
and \eqref{Eq:D02}.

We perform the calculations in the following way. First the system is 
stabilized in the thermal equilibrium. Then at time $t=t_0$ the external 
heat source and sink \eqref{Eq:D01} or \eqref{Eq:D02} is switched on 
where we chose equal values $Q_1=Q_2=Q$. Then after a time interval 
$\delta t=2.0\,\mathrm{s}$ all oscillations are relaxed and the system 
reaches a steady state. The local heat current 
$\mathbf{Q}(\mathbf{r},t)=Q\,\mathbf{e}_z$ will be homogenous in space, 
constant in time, and directed vertically along the $z$ axis.

The boundary conditions at $z_1$ and $z_2$ are important for the stability of 
the iterations. There should be no boundaries at all. This means we need periodic 
boundary conditions. The system can be made periodic in the following way.
We mirror the cell at one of the boundaries. Then we obtain a periodic structure 
of length $2L$. Furthermore, for the discretization the delta functions in 
Eqs.\ \eqref{Eq:D01} and \eqref{Eq:D02} must be replaced by smooth peaks of a 
small width $\delta z$. We choose $\delta z=3\,\Delta z$ which is a few 
discretization lengths. From the heat source at $z=z_1$ the heat current $Q_1$ 
will flow away in \emph{both} directions, where on the other hand a heat current 
$Q_2$ will flow from \emph{both} directions to the heat sink at $z_2$. This fact 
explains the factor $2$ in Eqs.\ \eqref{Eq:D01} and \eqref{Eq:D02}.

\subsection{Dimensionless renormalized quantities}
\label{Sec:04A}
\begin{figure}
\includegraphics[width=\linewidth,clip]{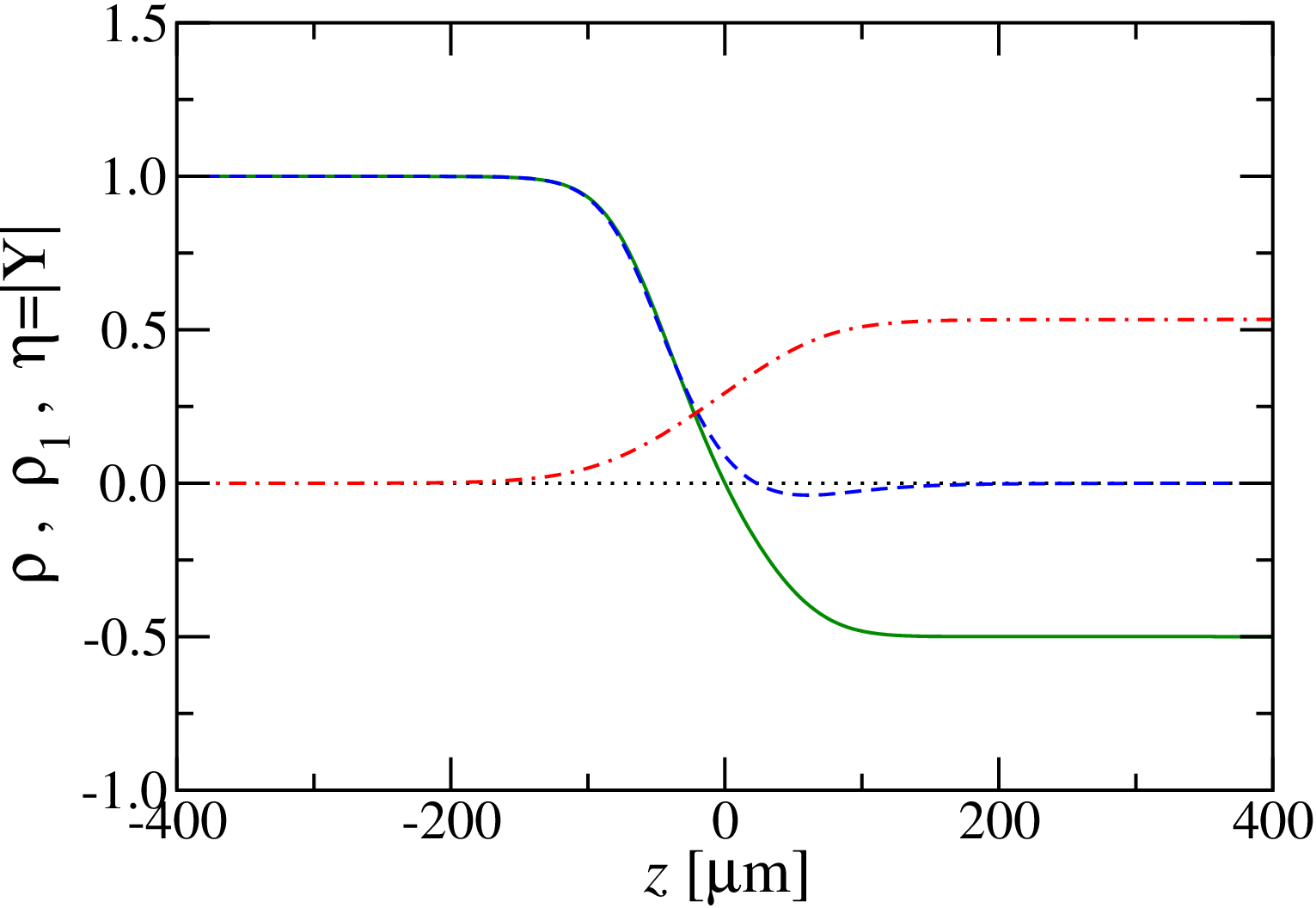}
\caption{(Color online) The dimensionless renormalized temperature parameters 
$\rho$ (green solid line), $\rho_1$ (blue dashed line), and the modulus of the 
order parameter $\eta=|Y|$ (red dash-dotted line) for the superfluid/normal-fluid
interface.}
\label{Fig:01}
\end{figure}

The direct results of the numerical calculation are the dimensionless renormalized 
temperature parameters $\Delta\rho$, $\rho$, $\rho_1$, and the dimensionless 
renormalized order parameter $Y$ as functions of the altitude coordinate $z$ and 
the time $t$. In Fig.\ \ref{Fig:01} the results are shown for the 
superfluid/normal-fluid interface of liquid $^4$He at zero heat current $Q=0$ in 
thermal equilibrium. The interface is induced by the gravitational acceleration 
$g=9.81\,\mathrm{m/s^2}$ on earth. Since in thermal equilibrium the temperature 
is constant we may choose it equal to the reference temperature so that $T=T_0$. 
Hence Eq.\ \eqref{Eq:C27} implies $\Delta\rho=0$. This is a trivial result which 
is shown by the black dotted line. The parameter $\rho$ is related to the temperature 
difference $T-T_\lambda$ by \eqref{Eq:C28} and shown as green solid line. The 
modified temperature parameter $\rho_1$ is defined in \eqref{Eq:C36} and shown as 
blue dashed line. Finally, the modulus of the dimensionless renormalized order 
parameter $\eta=|Y|$ is shown as red dash-dotted line.

In Fig.\ \ref{Fig:01} we observe three different regions. For low altitudes 
$z\lesssim -100\,\mathrm{\mu m}$ we find the asymptotic values $\rho\to 1$, 
$\rho_1\to 1$, and $\eta=|Y|\to 0$. Hence, in this region the $^4$He is normal fluid. 
We recover the related flow parameter condition $\rho=1$ of Ref.\ \onlinecite{Do85} 
in the asymptotic limit $z\to-\infty$. For high altitudes 
$z\gtrsim +100\,\mathrm{\mu m}$ we find the asymptotic values $\rho\to -1/2$, 
$\rho_1\to 0$, and $\eta=|Y|\to (A_d/8u)^{1/2}$ where $A_d=1/4\pi$ for $d=3$.
Hence, in this latter region the order parameter is nonzero and the $^4$He is 
superfluid. Again, we recover the related flow parameter condition $\rho=-1/2$ of 
Ref.\ \onlinecite{Do85} in the asymptotic limit $z\to+\infty$. The third region 
is the interface region $-100\,\mathrm{\mu m}\lesssim z \lesssim +100\,\mathrm{\mu m}$. 
Here the curves interpolate between the asymptotic values. We clearly see that the 
interface induced by gravity has a thickness of about 
$\Delta z_{I,g}\approx 200\,\mathrm{\mu m}$.

\begin{figure}
\includegraphics[width=\linewidth,clip]{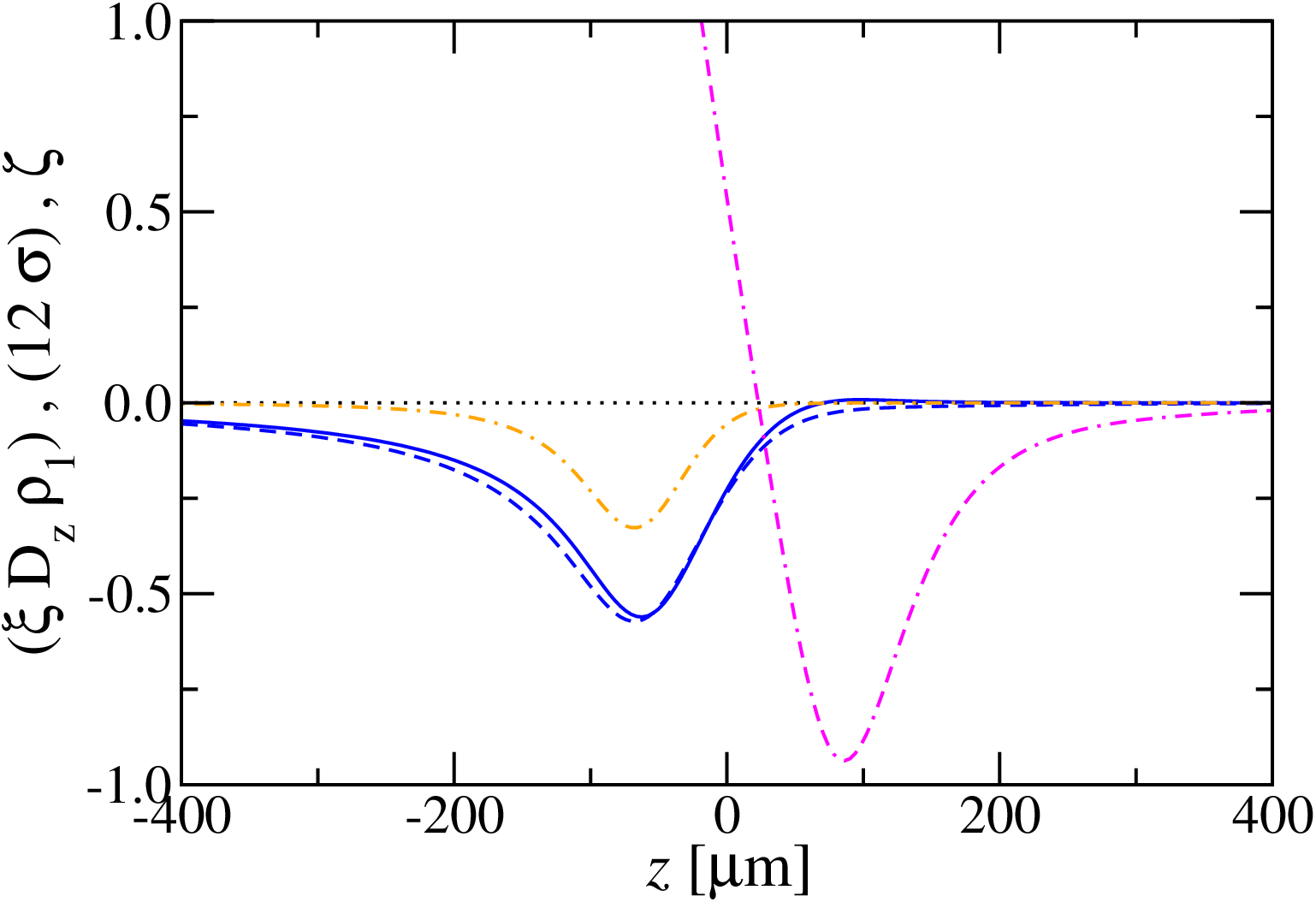}
\caption{(Color online) The blue lines show the dimensionless covariant 
derivative $\xi D_z\rho_1$ calculated in two ways: by numerical differentiation 
(blue solid line) and by formula \eqref{Eq:C68} (blue dashed line). Furthermore 
the parameter $\sigma$ defined in \eqref{Eq:C85} and multiplied by a factor $12$ 
is shown as orange dash-dotted line. Finally, the argument of the function 
\eqref{Eq:C50} $\bar\zeta$ is shown as magenta double-dash-dotted line.}
\label{Fig:02}
\end{figure}

Since the system is constant with respect to the horizontal coordinates $x$, $y$, 
and with respect to the time $t$, the covariant derivatives of the dimensionless 
renormalized quantities are nonzero only for the altitude coordinate $z$. In most 
cases these covariant derivatives are calculated by numerical differentiation 
using the formulas derived in section \ref{Sec:03E}. An exception is 
$\xi D_z \rho_1$ which is expressed in terms of other covariant derivatives by 
formula \eqref{Eq:C68}. The result is shown in Fig.\ \ref{Fig:02} by the blue 
dashed line. Alternatively, we apply Eq.\ \eqref{Eq:C67} to the modified 
temperature parameter $\rho_1$ and calculate the covariant derivative directly by 
numerical differentiation. This latter procedure is not correct in the interface 
region where the renormalization factors depend on the altitude coordinate because 
$\rho_1$ is not renormalized as $\rho$. Nevertheless, in Fig.\ \ref{Fig:02} the 
result is shown by the blue solid line. Surprisingly, the two blue lines, the solid 
one and the dashed one, are close to each other. Hence, Eq.\ \eqref{Eq:C67} is not 
that bad for calculating the covariant derivative $\xi D_z \rho_1$.

The blue lines in Fig.\ \ref{Fig:02} represent the covariant derivative of the 
blue dashed line in Fig.\ \ref{Fig:01}. However, the latter line represents 
$\rho_1=\rho_1(z)$ and shows a negative minimum value $\rho_{1,\mathrm{min}}<0$ 
at the position $z_\mathrm{min}\approx 60\,\mathrm{\mu m}$. Consequently for 
$\xi D_z \rho_1$ we expect a zero at this position related to a sign change. 
In Fig.\ \ref{Fig:02} the solid blue line does show this zero and sign change 
but the dashed blue line does not. In this way, the apparently incorrect formula 
\eqref{Eq:C67} for $\rho_1$ appears to be more realistic than the generic 
formula \eqref{Eq:C68}.

The existence of the sign change is supported by the following argument. In a small 
$z$ interval close to the interface we may modify the renormalization-group theory 
by choosing a constant flow parameter $\tau$. In this case the covariant derivatives 
reduce to the partial derivatives so that Eqs.\ \eqref{Eq:C67} and \eqref{Eq:C68} 
would yield identical results for $\xi D_z \rho_1=\xi \partial_z \rho_1$ and the 
two blue lines in Fig.\ \ref{Fig:02} would collapse to a single line. As a result 
the sign change would be found at $z_\mathrm{min}$ if we evaluate the partial 
derivative explicitly by differentiation of the blue dashed line in Fig.\ \ref{Fig:01}. 

However, the sign change of the solid blue line in Fig.\ \ref{Fig:02} would have 
a dramatic consequence for the numerical procedure when calculating $\sigma$ 
and the amplitudes $A$ and $A_1$. In thermal equilibrium we have $\Delta\rho=0$ 
so that Eq.\ \eqref{Eq:C85} reduces to $\sigma= -(\xi D_z \rho_1)^2/12$. 
Consequently, $\sigma$ will be negative everywhere except at a point close to 
$z_\mathrm{min}$. At this point we have $\sigma=0$ so that the formulas for the 
amplitudes $A$ and $A_1$ reduce to \eqref{Eq:C40} and \eqref{Eq:C41}, respectively. 
However, close to the minimum position $z_\mathrm{min}$ the modified temperature 
parameter $\rho_1$ is negative which implies an imaginary result for 
$\rho_1^{-\varepsilon/2}$ in Eqs.\ \eqref{Eq:C40} and \eqref{Eq:C41} where 
$\varepsilon=4-d=1$ for $d=3$. Hence, the amplitudes $A$ and $A_1$ are not well 
defined if the solid line in Fig.\ \ref{Fig:02} and the formula \eqref{Eq:C67} is used.

The problem arises due to the fact that we evaluate the Green function \eqref{Eq:B22},
the condensate density $n_\mathrm{s}$, the superfluid current $\mathbf{J}_\mathrm{s}$, 
and hence the amplitudes $A$ and $A_1$ in an approximation where only the covariant 
gradients of the temperature parameters $\rho_1$ and $\Delta \rho$ are taken into 
account. If we could do the calculation for the full space dependence all these 
quantities would be well defined. In the unpublished work \cite{HN08} we extended 
the calculation by including also the curvatures of the temperature parameters. 
While the problem at $z_\mathrm{min}$ was abolished, the calculation was much more 
complicated and restricted to the thermal equilibrium at zero heat current. Moreover, 
other mathematical difficulties appeared. Hence, this more sophisticated calculation 
could not be realized in practice for our purpose.

However, our fortune is a small inaccuracy of the approximation in our numerical 
calculation in practice which implies that the blue dashed line in Fig.\ \ref{Fig:02} 
is completely negative and does not show a zero and a sign change for the covariant 
derivative $\xi D_z \rho_1$. The related parameter $\sigma= -(\xi D_z \rho_1)^2/12$ 
is shown in Fig.\ \ref{Fig:02} as orange dash-dotted line where it has been enhanced 
by a factor $12$. Clearly, this curve is negative and never zero for all altitudes 
$z$ in the interface region. For this reason we can apply our formulas \eqref{Eq:C51} 
and \eqref{Eq:C52} for the amplitudes $A$ and $A_1$ without a problem if we use the 
generic formula \eqref{Eq:C68} for the dimensionless gradient $\xi D_z \rho_1$. We 
obtain smooth and stable results which are within the accuracy of our approximation.

In order to evaluate the amplitudes $A$ and $A_1$ we need the function 
\eqref{Eq:C50} and its argument $\bar\zeta=\rho_1 / (-\sigma)^{1/3}$. Consequently, 
from the dashed blue line in Fig.\ \ref{Fig:01} and the orange dash-dotted line in 
Fig.\ \ref{Fig:02} we obtain the dimensionless variable $\bar\zeta$ as a function 
of the altitude coordinate $z$ which is shown in Fig.\ \ref{Fig:02} by the magenta 
double-dash-dotted line. In the normal-fluid region for $z<0$ the variable 
$\bar\zeta$ increases quickly for decreasing altitude $z$. Consequently, in this 
case the asymptotic formula \eqref{Eq:C53} can be used so that the amplitudes 
$A$ and $A_1$ reduce to the simple formulas \eqref{Eq:C40} and \eqref{Eq:C41} of 
the normal-fluid equilibrium state. In the superfluid region near the interface 
the variable $\bar\zeta$ is negative. However, it is bounded from below by the 
value $-1$. Consequently, the asymptotic formula \eqref{Eq:C54} is not needed. 
This means that the variable $\bar\zeta$ never comes in the large negative region 
where the function \eqref{Eq:C50} oscillates and possesses a significant imaginary 
part. This observation is very important for the consistency of our theory because 
the oscillations would be unphysical and the imaginary part would be related to 
an instability.

\subsection{Temperature profiles}
\label{Sec:04B}
Until now, the calculations are restricted to the thermal equilibrium at zero 
heat current $Q=0$. Here the phase of the order parameter $Y=\eta\,e^{i\varphi}$ 
is constant, so that we can chose $\varphi=0$. We have extended our numerical 
calculations to small nonzero heat currents $Q$ in the interval 
$-70\,\mathrm{nW/cm^2}\leq Q \leq +160\,\mathrm{nW/cm^2}$. In this latter case 
the phase of the order parameter $\varphi=\varphi(z,t)$ will be a nontrivial 
function of the altitude coordinate $z$ and the time $t$. A positive heat current 
$Q>0$ means a heat flow $\mathbf{Q}=Q\,\mathbf{e}_z$ in the $z$ direction which means 
that the heat current flows upward from bottom to top. The original experiment 
by Duncan \emph{et al.}\ \cite{DAS88} and succeeding experiments investigating 
the superfluid/normal-fluid interface induced by a heat current $Q$ were 
performed in this configuration. On the other hand a negative heat current $Q<0$ 
means a downward heat flow from top to bottom. This latter configuration was 
investigated much later in the experiment by Moeur \emph{et al.}\ \cite{Du97}. 

In the nonequilibrium system with a nonzero heat flow the dimensionless 
renormalized temperature parameter $\Delta\rho$ will be nonzero. Once the local 
space and time dependent RG flow parameter $\tau=\tau(z,t)$ is known, the space 
and time dependent temperature profile $T=T(z,t)$ is calculated from $\Delta\rho$ 
by Eq.\ \eqref{Eq:C27}. Furthermore, the local space and time dependent heat 
current $Q=Q(z,t)$ is calculated from the dimensionless renormalized heat 
current $\tilde q$ by Eq.\ \eqref{Eq:D03}. After a time difference of about 
$\delta t=2\,\mathrm{s}$ the system will relax in a stationary state where all 
quantities are constant in time. If in Eq.\ \eqref{Eq:D01} the source and sink 
parameters $Q_1=Q_2=Q$ are chosen, a vertical heat current $Q$ will be found in 
the whole system which is constant in the space variable $z$. Consequently, the 
different temperature profiles we obtain in our numerical calculations can be 
labeled by this constant heat current.

\begin{figure}
\includegraphics[width=\linewidth,clip]{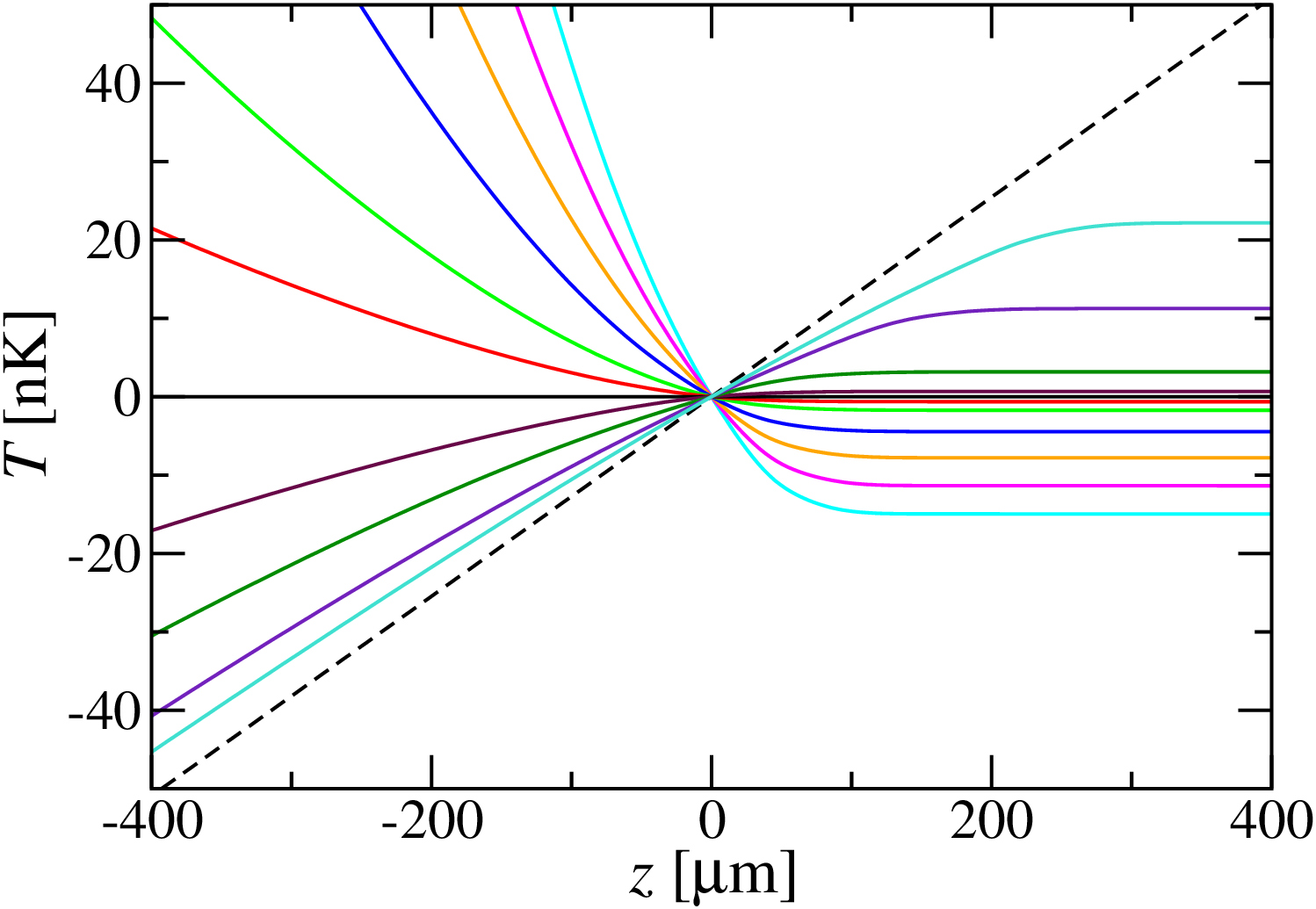}
\caption{(Color online) The temperature profiles $T(z)$ of the 
superfluid/normal-fluid interface of liquid $^4$He in gravity are shown for several 
heat currents $Q$ as colored solid lines. The solid lines on the left-hand (normal-fluid)
side are ordered from top to bottom with respect to decreasing heat currents $Q=160$, 
$130$, $100$, $70$, $40$, $20$, $0$, $-20$, $-40$, $-60$, $-70\,\mathrm{nW/cm^2}$.
On the right-hand (superfluid) side they are ordered from bottom to top.
The horizontal black solid line represents the temperature in thermal equilibrium 
for $Q=0$. The straight black dashed line represents the superfluid transition 
temperature $T_\lambda(z)$.}
\label{Fig:03}
\end{figure}

Our numerical results are shown in Fig.\ \ref{Fig:03}. The temperature profile 
$T(z)$ is shown by the colored solid lines for several values of the heat current 
$Q$ which are specified in the caption of the figure. On the other hand, the 
superfluid transition temperature $T_\lambda(z)$ as a function of the altitude 
coordinate $z$ is shown by the straight black dashed line. The slope of this 
latter line is the effect of the gravity on earth.

The altitude $z_0$ at which the temperature profiles $T(z)$ and $T_\lambda(z)$ 
intersect each other so that $T(z_0)= T_\lambda(z_0)$ may be viewed as a reference 
altitude to specify the position of the superfluid/normal-fluid interface. We 
realize that the system is translation invariant in the sense that we can move 
the curves parallel along the straight dashed line. Thus, for convenience and 
simplicity we select a coordinate system so that all curves intersect at the same 
altitude $z=z_0=0$. This choice is no physical restriction and has been applied 
in Fig.\ \ref{Fig:03}.

On the left hand side for low altitudes $z\lesssim -100\,\mathrm{\mu m}$ the 
system is normal fluid. Here the heat transport equation $Q=-\lambda_T \partial_z T$ 
implies that the temperature gradient $\partial_z T$ is negative for positive 
heat currents $Q$ and positive for negative heat currents. The values of the 
gradients are considerably large. On the right hand side for high altitudes 
$z\gtrsim +100\,\mathrm{\mu m}$ the system is superfluid. Here the heat is 
transported convectively following the two-fluid model so that the temperatures 
$T(z)$ are nearly constant and the gradients are nearly zero. The intermediate 
region $-100\,\mathrm{\mu m} \lesssim z\lesssim +100\,\mathrm{\mu m}$ is the 
superfluid/normal-fluid interface. Here the temperature profiles interpolate 
the two outer regions.

For positive heat currents $Q>0$ (heat flow upward) the slope of the temperature 
curve $T(z)$ increases without a limit on the normal-fluid side for $z\to -\infty$. 
However, for negative heat currents $Q<0$ (heat flow downward) the slope increases 
up to a limiting value which is the slope of $T_\lambda(z)$ so that in the limit 
$z\to -\infty$ the temperature profile $T(z)$ approaches a straight line parallel 
to the straight dashed line $T_\lambda(z)$. This latter fact is clearly observed 
in the lower left part of Fig.\ \ref{Fig:03}. It represents the 
\emph{self-organized critical state} predicted by Onuki \cite{On87} 
and discovered in the experiment by Moeur \emph{et al.}\ \cite{Du97}.

While Figs.\ \ref{Fig:01} and \ref{Fig:02} are calculated for the thermal 
equilibrium at zero heat current $Q=0$, we have calculated the related curves also 
for the nonequilibrium state at the nonzero heat currents of Fig.\ \ref{Fig:03}. 
Most curves do not change very much, the characteristic forms remain qualitatively. 
An exception is the parameter $\sigma$ defined in Eq.\ \eqref{Eq:C85} and shown as 
orange dash-dotted line in Fig.\ \ref{Fig:02}. This parameter is negative in the 
whole system only for small heat currents in the interval 
$-10\,\mathrm{nW/cm^2}\lesssim Q \lesssim +20\,\mathrm{nW/cm^2}$. For larger heat 
currents outside this interval the parameter $\sigma$ will change the sign from 
negative to positive at specific altitudes $z$. For even larger negative heat 
currents $Q\lesssim -20\,\mathrm{nW/cm^2}$ and even larger positive heat currents 
$Q\gtrsim +40\,\mathrm{nW/cm^2}$ the parameter $\sigma$ is positive in the whole 
system.

\subsection{Order parameter}
\label{Sec:04C}
The order parameter in physical units $\langle\psi\rangle$ is calculated from the 
dimensionless renormalized order parameter $Y$ via the renormalization formulas 
\eqref{Eq:C01} and \eqref{Eq:C31}. Putting these equations together and replacing 
$\mu\to\xi^{-1}$ we obtain
\begin{equation}
\langle\psi\rangle = Z_\phi^{1/2}\, Y \, \xi^{-(d-2)/2} \ .
\label{Eq:D04}
\end{equation}
Integrating the defining equation \eqref{Eq:C56} for the zeta function $\zeta_\phi$, 
we obtain an integral representation for the renormalization factor
\begin{equation}
Z_\phi = \exp\Bigl\{ \int_\mu^\infty \zeta_\phi \, \frac{d\mu^\prime}{\mu^\prime} \Bigr\} 
= \exp\Bigl\{ - \int_\tau^\infty \nu\eta \, \frac{d\tau^\prime}{\tau^\prime} \Bigr\} \ .
\label{Eq:D05}
\end{equation}
The second equality sign is implied by the flow-parameter transformation \eqref{Eq:C60} 
together with the running exponents $\nu$ and $\eta$, defined in \eqref{Eq:C70} and 
\eqref{Eq:C71}. The upper infinite integration boundaries guarantee $Z_\phi=1$ in the 
limits $\mu\to\infty$ and $\tau\to\infty$ which represent the mean-field or Gaussian 
fix point of the RG flow. If we use the correlation length $\xi=\xi_0\, \tau^{-\nu}$ 
we obtain the asymptotic formula for the order parameter 
$\langle\psi\rangle\sim \tau^{\nu(d-2+\eta)/2}=\tau^\beta$ with the correct critical 
exponent $\beta$ defined in \eqref{Eq:C72}.

\begin{figure}
\includegraphics[width=\linewidth,clip]{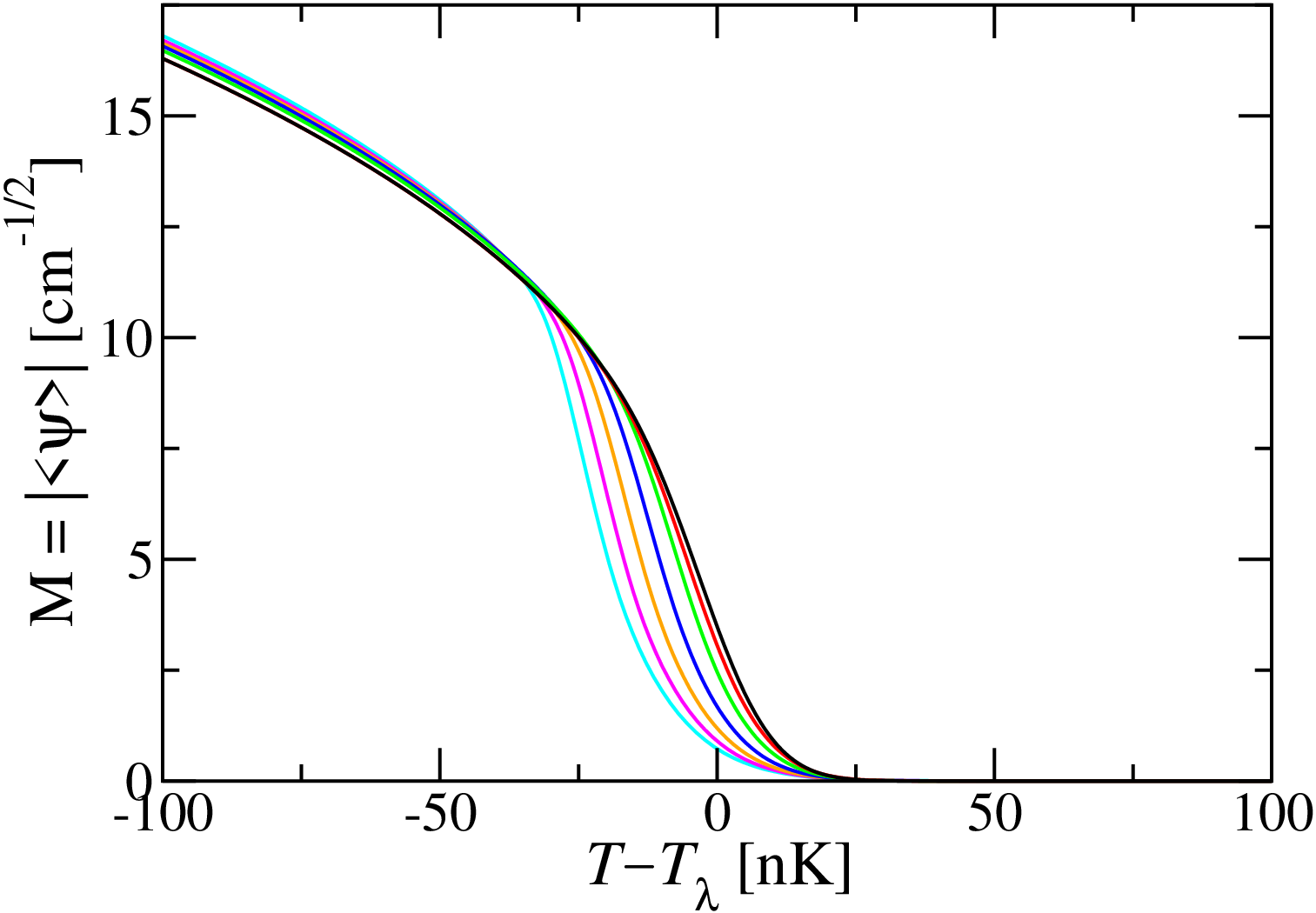}
\caption{(Color online) The modulus of the order parameter $M=|\langle\psi\rangle|$ 
as a function of the temperature difference $T-T_\lambda$ for the superfluid/normal-fluid 
interface in gravity. The colored solid lines from left to right represent the heat 
currents $Q=160$, $130$, $100$, $70$, $40$, $20$, $0\,\mathrm{nW/cm^2}$.}
\label{Fig:04}
\end{figure}

Eqs.\ \eqref{Eq:D04} and \eqref{Eq:D05} are suited for a numerical calculation once 
the dimensionless renormalized order parameter $Y$, the RG flow parameter $\tau$, 
and the running exponents \eqref{Eq:C70}-\eqref{Eq:C72} are known. We have calculated 
the order parameter $\langle\psi\rangle= M\, e^{i\varphi}$ in the stationary state 
for all those heat currents $Q$ for which we have calculated the temperature profiles 
in the previous subsection. We obtain the modulus $M$ and the phase $\varphi$ of 
the order parameter. Our results for the modulus $M$ are shown in Fig.\ \ref{Fig:04} 
for positive heat currents $Q\geq 0$ (heat flow upward). The colors of the solid lines 
correspond to those in Fig.\ \ref{Fig:03}. Here and in the following figures we omit 
the lines for negative heat currents $Q<0$ (heat flow downward) because they make the 
figures complicated and involved but do now show new physics.

Close to criticality $T=T_\lambda$ the curves are smooth. This is an effect of 
gravity and related to the superfluid/normal-fluid interface. The width of the 
smooth region is $\Delta T_{I,g}=25\,\mathrm{nK}$ which corresponds to the 
thickness of the interface $\Delta z_{I,g}=200\,\mathrm{\mu m}$. The ratio is 
approximately the gradient of the superfluid transition temperature, i.e.\ 
$\Delta T_{I,g}/\Delta z_{I,g} \approx \partial T_\lambda/\partial z 
= 1.273\,\mathrm{\mu K/cm}$. For increasing heat currents $Q$ the smooth curves 
are shifted to the left to lower temperatures. This fact is related to the 
depression of the superfluid transition to lower temperatures by a heat current 
which has been observed and investigated in the experiment by Duncan, Ahlers, 
and Steinberg \cite{DAS88}.

Away from criticality for lower temperatures $T-T_\lambda\lesssim -30\,\mathrm{nK}$ 
the curves approach asymptotically a single line which corresponds to the 
singular order parameter $M=|\langle\psi\rangle|\sim (T_\lambda - T)^\beta$ for 
$T<T_\lambda$ in thermal equilibrium and zero gravity. In Fig.\ \ref{Fig:04} the 
asymptotic curves do not fall perfectly on a single line. This observation is 
a numerical error in our calculation. In order to stabilize the numerical iterations 
we must add an imaginary part to the parameter $\sigma$ defined in \eqref{Eq:C85}. 
This imaginary part increases with increasing heat current $Q$ and influences 
slightly the curves on the superfluid side.

The physical units $\mathrm{cm}^{-1/2}$ of the order parameter arising from 
the formula \eqref{Eq:D04} for $d=3$ dimensions appear to be artificial and 
unphysical. However, since the order parameter can not be observed in physical 
experiments, this artifact is not important and no matter of concern.

The phase of the order parameter $\varphi$ is dimensionless. Its gradient is 
related to the superfluid velocity $\mathbf{v}_\mathrm{s} = (\hbar/m_4)\nabla\varphi$. 
For nonzero heat currents $Q$ we find nontrivial results for the superfluid velocity 
$v_\mathrm{s}$. If we approach the interface from the superfluid side, $v_\mathrm{s}$ 
increases monotonically. However, on the normal-fluid side, the phase $\varphi$ 
and the superfluid velocity $v_\mathrm{s}$ are irrelevant because the modulus $M$ 
approaches zero.

\subsection{Correlation length}
\label{Sec:04D}
\begin{figure}
\includegraphics[width=\linewidth,clip]{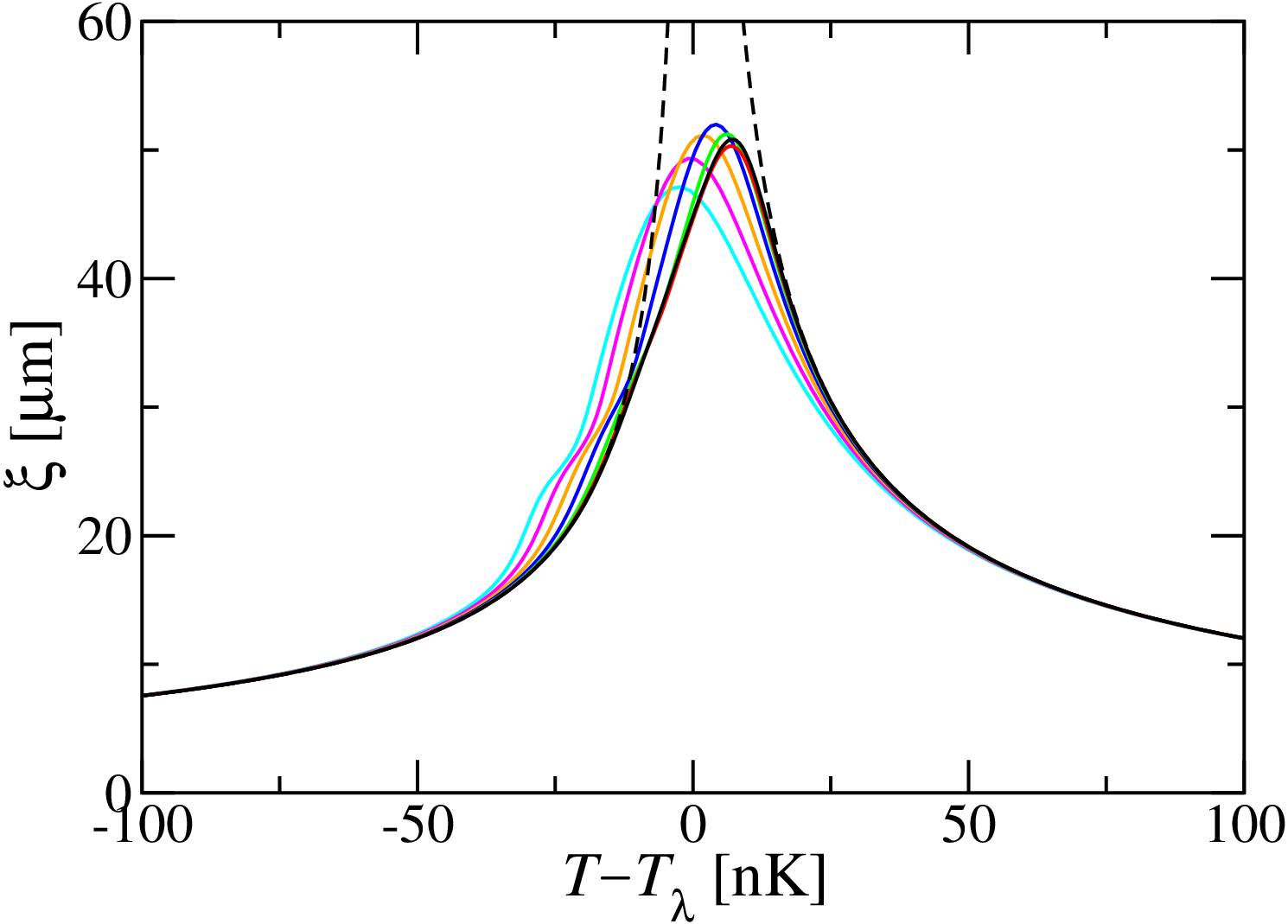}
\caption{(Color online) The correlation length $\xi$ as a function of the 
temperature difference $T-T_\lambda$ for the superfluid/normal-fluid interface 
in gravity. The colored solid lines from left to right represent the heat currents 
$Q=160$, $130$, $100$, $70$, $40$, $20$, $0\,\mathrm{nW/cm^2}$. As a reference 
the black dashed line shows the singular correlation length in thermal equilibrium 
and zero gravity.}
\label{Fig:05}
\end{figure}

The correlation length $\xi$ has been calculated by Schloms and Dohm \cite{SD89} 
in thermal equilibrium and zero gravity. In the renormalized perturbation 
theory up to two-loop order they obtain $\xi^{-2}=\mu^2 A_\xi$ with an amplitude 
function $A_\xi=1+\mathcal{O}(u^2)$. However, since our Hartree approximation is 
first order in $u$ we may approximate $A_\xi\approx 1$, so that the correlation 
length is just $\xi=\mu^{-1}$. This quantity is provided by our numerical calculation. 
Our results are shown in Fig.\ \ref{Fig:05} by the colored solid lines for the same 
positive heat currents as in the previous figures. In the interface region close 
to criticality $T=T_\lambda$ the colored solid curves are smooth. The correlation 
length has a maximum value $\xi_g\approx 50\,\mathrm{\mu m}$ which is implied by 
the gravity acceleration $g=9.81\,\mathrm{m/s^2}$ on earth. The effect of a small 
nonzero heat current $Q$ is weak. For increasing heat currents $Q$ the position 
of the maximum of the correlation length is shifted slightly to lower temperatures. 
We note that our maximum correlation length $\xi_g$ is of the same order of magnitude 
as the characteristic length $l_g=67\,\mathrm{\mu m}$ which was used by Ginzburg 
and Sobyanin \cite{GS76} within their $\psi$ theory.

From Fig.\ \ref{Fig:01} we have inferred the interface thickness 
$\Delta z_{I,g}=200\,\mathrm{\mu m}$. Thus we calculate the ratio 
$\Delta z_{I,g} / \xi_g \approx 4$ which means that the interface thickness 
is four times the maximum of the correlation length. While in a nonequilibrium 
and/or gravity environment the correlation length $\xi$ is finite and a smooth 
function, in equilibrium and zero gravity it shows the well known singular 
behavior $\xi\sim |T-T_\lambda|^{-\nu}$ near criticality for $T\to T_\lambda$
with an exponent $\nu=0.671$. This latter singular correlation length is 
shown by the black dashed line which diverges at $T=T_\lambda$. Far away from 
criticality which means far away from the interface all solid lines converge 
to a single line which is identical with the black dashed line. Thus, far away 
from the interface the gravity $g$ and the heat current $Q$ do not have an 
influence on the correlation length $\xi$. Finally, here we do not see an influence 
of the imaginary part of the parameter $\sigma$ we introduce in our calculation 
in order to stabilize the numerical iterations.

\subsection{Specific heat}
\label{Sec:04E}
There are two possibilities to calculate the specific heat. First, we may calculate 
the entropy $S$ within our renormalization-group theory and then calculate the 
derivative $C_X=T ( \partial S / \partial T )_X$ numerically where any quantity $X$ 
may be kept constant. This has been done in our previous paper \cite{Ha99a} where 
$X=Q$ or $X=\nabla T$. The entropy $S$ is given by Eqs.\ (8.10) or (8.12) of Ref.\ 
\onlinecite{Ha99a}. Secondly, we calculate the specific heat directly by 
$C=k_\mathrm{B}\chi_0 Z_m C_\mathrm{ren}$ where the renormalized specific heat 
$C_\mathrm{ren}$ is defined in \eqref{Eq:C22} and the renormalization factor 
$\chi_0 Z_m$ is defined implicitly in \eqref{Eq:C30}. Thus, we obtain
\begin{equation}
C_X = k_\mathrm{B} \frac{A_d}{4\tau^2 \xi^d} \, \Bigl\{ \frac{1}{\gamma^2}
+ \frac{1}{2u} \Bigl[ 1 - \Bigl( \frac{\partial\rho_1}{\partial\rho} \Bigr)_X 
\Bigr] \Bigr\} \ ,
\label{Eq:D06}
\end{equation}
a formula which should be compared with the entropy (8.10) in Ref.\ \onlinecite{Ha99a}.
The formula can be simplified if we use the asymptotic formulas for the correlation 
length $\xi=\xi_0\, \tau^{-\nu}$ and for the coupling parameter 
$\gamma^{-2}=(4\nu/\alpha)(1-b \tau^\alpha)$ where $\nu=0.671$ and 
$\alpha=2-d\nu=-0.013$ are critical exponents and $b$ is a known constant. 
As a result we obtain the specific heat
\begin{equation}
C_X = B + \tilde A \{ (4\nu/\alpha) + F_X[u] \} \tau^{-\alpha}
\label{Eq:D07}
\end{equation}
together with the amplitude
\begin{equation}
F_X[u] = \frac{1}{2u} \Bigl[ 1 - \Bigl( \frac{\partial\rho_1}{\partial\rho} 
\Bigr)_X \Bigr] \ .
\label{Eq:D08}
\end{equation}
This formula should be compared with the entropy (8.12) in our previous paper 
\cite{Ha99a} together with Eqs.\ (8.13)-(8.16). Here $\tilde A$ and $B$ are 
nonuniversal constants which can be expressed in the forms 
$\tilde A= k_\mathrm{B} A_d / 4 \xi_0^d$ and $B= \tilde A (-4\nu/\alpha) b$.
Alternatively, these constants can be obtained by fitting the formula to the 
experimental data for liquid $^4$He in a micro gravity environment in space 
\cite{Li96,Li03}. In this way we obtain $\tilde A= 2.22\,\mathrm{J/mol\, K}$ and 
$B=456\,\mathrm{J/mol\, K}$ where the constants are multiplied additionally 
by the molar volume of liquid $^4$He at saturated vapor pressure \cite{Al85} 
$V_\lambda=27.38\,\mathrm{cm^3/mol}$.

The amplitude $F_X[u]$ can be compared directly with the amplitudes $F_\pm[u]$ 
of Dohm \cite{Do85}, if we consider the asymptotic limits far away from the 
interface. The temperature parameters $\rho$ and $\rho_1$ are related to each 
other by \eqref{Eq:C36}. In the normal-fluid region far away from the interface 
the renormalized order parameter is $Y=0$ and the amplitudes $A$ and $A_1$ are 
given by \eqref{Eq:C40} and \eqref{Eq:C41}. The partial derivative can be performed 
easily so that we obtain $\partial\rho/\partial\rho_1 = 1 + 8u\, A_1= 1 - 4u$ 
which does not depend on the variable $X$ that is kept constant. Thus we obtain 
the amplitude
\begin{equation}
F_+[u] = (2u)^{-1} [ 1 - 1 / ( 1 - 4u ) ] = - 2 + \mathcal{O}(u)  \ .
\label{Eq:D09}
\end{equation}
In the superfluid region $\rho_1$ approaches $0$ more rapidly than $\rho$ 
approaches $-1/2$. Consequently, in the superfluid region far away from the 
interface the partial derivative is $\partial\rho_1/\partial\rho=0$ which again 
does not depend on the variable $X$ that is kept constant. Thus we obtain 
the amplitude $F_-[u] = (2u)^{-1}$. If we compare our results for $F_\pm[u]$ 
with those of Dohm \cite{Do85} we find agreement for the leading terms in 
powers of $u$ in both cases $+$ and $-$, respectively.

We have calculated the specific heat numerically with both methods described 
above using the entropy formula (8.12) of our previous paper \cite{Ha99a} and 
the specific-heat formula \eqref{Eq:D07} of the present paper. The results 
agree with each other within the accuracy of our Hartree approximation which 
is a self-consistent one-loop approximation combined with the 
renormalization-group theory. This agreement is a test for the validity and 
the accuracy of our method presented in this paper. While in the previous 
paper we have used the first method, in this paper we prefer the second method,
i.e.\ formula \eqref{Eq:D07} together with \eqref{Eq:D08}. The reason is that 
in the present calculation the second method provides curves looking smoother 
and more nice. 

\begin{figure}
\includegraphics[width=\linewidth,clip]{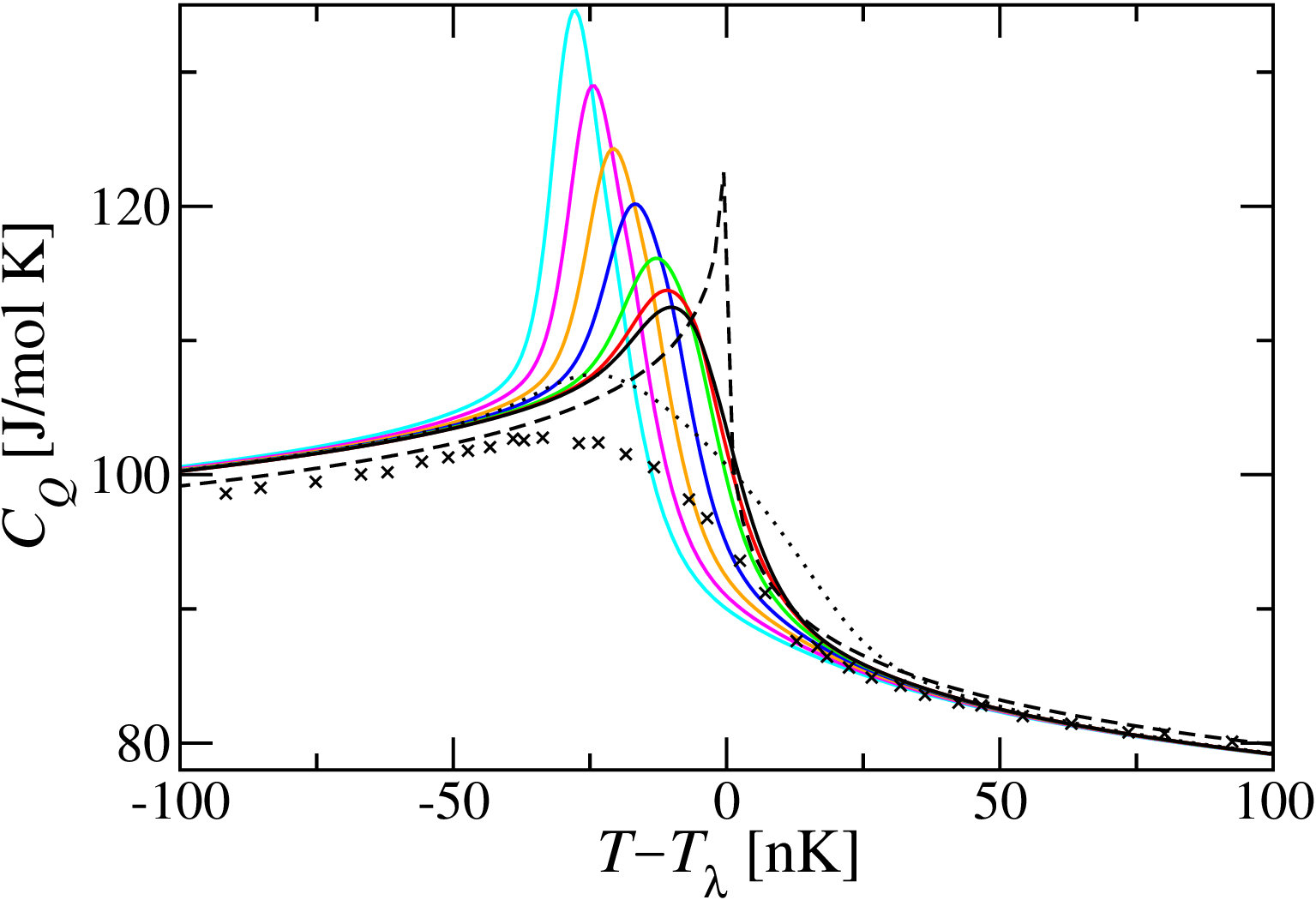}
\caption{(Color online) The specific heat $C_Q$ as a function of the 
temperature difference $T-T_\lambda$ for the superfluid/normal-fluid interface 
in gravity. The colored solid lines from left to right represent the heat currents 
$Q=160$, $130$, $100$, $70$, $40$, $20$, $0\,\mathrm{nW/cm^2}$. The black dashed 
line represents the singular specific heat in microgravity fitted to the data of 
the experiment by Lipa \emph{et al.}\ \cite{Li96,Li03}. The black crosses show the 
experimental data for zero heat current in gravity on earth by Lipa \cite{Li11} 
where the black dotted line is the related theoretical curve for the average 
specific heat $\bar C_Q$.}
\label{Fig:06}
\end{figure}

Our results are shown in Fig.\ \ref{Fig:06} by the colored solid lines for the 
same heat currents as in the previous figures. We have calculated the specific 
heat $C_Q$ where the heat current $Q$ and the gravity acceleration 
$g=9.81\,\mathrm{cm/s^2}$ are kept constant. Clearly, in the interface region 
near criticality we find smooth curves. The specific heat has a maximum slightly 
below the critical temperature. For increasing heat currents $Q$ this maximum is 
shifted to lower temperatures which is related to the depression of the superfluid 
transition temperature observed in the experiment by Duncan, Ahlers, and Steinberg 
\cite{DAS88}. Furthermore, the maximum of the specific heat is strongly enhanced 
for increasing heat currents $Q$. This enhancement is an effect of the constant 
heat current $Q$ when calculating the specific heat $C_Q$. It has been observed 
already in our previous paper \cite{Ha99a}, where $C_Q$ has be calculated for 
the much higher heat current $Q=42.9\,\mathrm{\mu W/cm^2}$ where gravity effects 
are negligible. The strong enhancement of the maximum is also compatible with 
experimental measurements of $C_Q$ by Harter \emph{et al.}\ \cite{HL00}.

Far away from criticality and the interface on both sides the colored solid curves 
converge to a single line, respectively. These single lines represent the asymptotic 
specific heat $C=B+(A_\pm/\alpha)|t|^{-\alpha}$ where $t=(T-T_\lambda)/T_\lambda$ 
is the reduced temperature and $\alpha$ is the critical exponent. On the 
normal-fluid side the single line is perfect. However, on the superfluid side it 
is slightly influenced by the imaginary part of the parameter $\sigma$ which we 
must add in our numerical calculation in order to have stable iterations. This 
fact is related to the similar observation in our results for the order parameter 
shown in Fig.\ \ref{Fig:04}.

The smooth colored solid lines in Fig.\ \ref{Fig:06} show that the critical 
singularity is rounded by the gravity $g$ and the heat current $Q$. The 
temperature scale for this rounding is $\Delta T_{g,I}=25\,\mathrm{nK}$ if 
gravity is the dominating effect. We have obtained this value from the thickness 
of the interface $\Delta z_{I,g}=200\,\mathrm{\mu m}$. Consequently, the asymptotic 
critical behavior of the specific heat and all other singular quantities can be 
observed only for temperatures $|T-T_\lambda|\gtrsim \Delta T_{g,I}=25\,\mathrm{nK}$
away from criticality. Hence, the gravity implies that on earth the critical point 
can never be reached. For this reason, experiments to measure the asymptotic behavior 
closer to the critical point must be performed in a micro-gravity environment in 
space.

Lipa \emph{et al.}\ \cite{Li96,Li03} have performed a space experiment which was 
called \emph{Lambda Point Experiment} (LPE) and which flew aboard the space shuttle 
Columbia (STS-52) in 1992. They obtained data for the specific heat up to 
$|T-T_\lambda|=1\,\mathrm{nK}$. They fitted an asymptotic formula to the data and 
determined the exponent $\alpha=-0.013$, the amplitudes $A_\pm$ and $B$ and some 
further parameters. The resulting fit curve is shown in Fig.\ \ref{Fig:06} by the 
black dashed line. This curve shows the typical lambda of the specific heat with 
a singularity at $T=T_\lambda$. Away from criticality on both sides for 
$|T-T_\lambda|\gtrsim 50\,\mathrm{nK}$ the solid lines and the dashed line come 
close to each other which demonstrates the agreement between theory and experiment. 
However, the agreement is not perfect. There remains a small discrepancy which is 
due to the amplitude ratio $A_+/A_-$ because our theory provides an approximate 
value for this amplitude ratio which can never be identical to the experimental 
value.

We note that in our calculation the specific heat is defined locally. It depends 
on the altitude $z$ so that $C_Q=C_Q(z)$. The rounding of the critical singularity 
is caused by the gradient of $T_\lambda(z)$ and by the heat current only. However, 
in experiments the $^4$He is in a cell of a finite extension. The cell is usually 
confined by two horizontal plates at altitudes $z_1$ and $z_2$, where the vertical 
height $L=z_2-z_1$ is small and the horizontal extensions are large. For this reason 
the experiment measures an average specific heat which is defined by the integral
$\bar{C}_Q=(z_2-z_1)^{-1} \int_{z_1}^{z_2} dz\, C_Q(z)$. This average process 
smooths the curve additionally. The maximum of the average specific heat 
$\bar{C}_Q$ will be broader than the maximum of the related local specific heat 
$C_Q$.

In order to minimize the averaging effects the cell height $L$ should be chosen 
as small as possible. However, it must be considerably larger than the maximum 
correlation length $\xi_g=50\,\mu m$ because for small $L$ finite size effects 
occur which again smooth and round the critical singularity. Consequently, for 
the cell height $L$ there will be an optimum range to obtain best measurements 
on earth. 

In Fig.\ \ref{Fig:06} the black crosses represent the experimental data of a 
measurement on earth at zero heat current performed by Lipa \cite{Li11}. In this 
case the cell height is $L=0.38\, mm$ which causes considerable average effects. 
The maximum of the experimental data is much broader than the maximum of the black 
solid line which represents the local specific heat $C_Q$ for $Q=0$. We have 
calculated the related average specific heat $\bar C_Q$ for $Q=0$ which is shown 
by the black dotted line. This latter curve shows a much broader maximum at 
criticality which agrees with the experimental data. Since the ratio $L/\xi_g=7.6$ 
is large, finite size effects are small. However, the Dirichlet boundary 
conditions of the order parameter at the cell walls imply that nevertheless the 
finite size effects cause a depression of the data which is clearly observed in 
Fig.\ \ref{Fig:06} because the experimental data (black crosses) are below the 
theoretical curve (black dotted line). Thus we conclude that the experimental 
data obtained in a measurement in gravity on earth agree with our theory.

\subsection{Thermal conductivity and resistivity}
\label{Sec:04F}
The thermal conductivity $\lambda_\mathrm{T}$ is defined locally by the heat 
transport equation $\mathbf{Q}=-\lambda_\mathrm{T} \nabla T$. Resolving this 
equation we obtain the thermal conductivity explicitly as 
$\lambda_\mathrm{T} = |\mathbf{Q}|/|\nabla T|$. Next we insert the renormalization 
equations for the heat current \eqref{Eq:D03} and for the temperature gradient 
\eqref{Eq:C65}. Thus, we obtain
\begin{equation}
\lambda_\mathrm{T} = \frac{g_0 k_\mathrm{B}}{\tau\,\xi^{d-2}} 
\,\frac{|\tilde{\mathbf{q}}|}{|\xi\mathbf{D}\Delta\rho|} \ .
\label{Eq:D10}
\end{equation}
The dimensionless heat current $\tilde{\mathbf{q}}$ and the dimensionless 
renormalized gradient $\xi\mathbf{D}\Delta\rho$ are variables in our numerical 
calculation. Hence Eq.\ \eqref{Eq:D10} is well suited for an explicit calculation 
of the thermal conductivity.

The dimensionless renormalized heat current is defined in \eqref{Eq:C82}.
Far away from the interface in the normal-fluid region the order parameter $Y$ 
and hence the last term of \eqref{Eq:C82} is zero. On the other hand, the amplitude 
$A_1$ in the first term of \eqref{Eq:C82} reduces to $A_1=-1/2$ following 
\eqref{Eq:C41}. Thus Eq.\ \eqref{Eq:D10} provides the thermal conductivity 
\begin{equation}
\lambda_\mathrm{T} = \frac{g_0 k_\mathrm{B}}{\tau\,\xi^{d-2}} 
\,\frac{A_d}{2\gamma F}\, \Bigl\{ 1 - \frac{f}{4} \Bigr\} \ .
\label{Eq:D11}
\end{equation}
in the normal-fluid region. This result is well known and agrees with the 
linear-response calculation of Dohm \cite{Do85}. Since our result is derived in 
Hartree approximation, the agreement is up to one-loop order.

Far away from the interface in the superfluid region the heat current $\mathbf{Q}$ 
is nonzero where the temperature gradient $\nabla T$ is zero. Hence the thermal 
conductivity $\lambda_\mathrm{T}$ is infinite. This fact is also seen in the formulas 
\eqref{Eq:C82} and \eqref{Eq:D10}. A zero dimensionless renormalized temperature 
gradient $\xi\mathbf{D}\Delta\rho=\mathbf{0}$ implies a zero first term in 
\eqref{Eq:C82} and a zero denominator in \eqref{Eq:D10}. On the other hand, the 
second term in \eqref{Eq:C82} is minus the dimensionless renormalized superfluid 
current which is nonzero. Consequently, Eq.\ \eqref{Eq:D10} provides an infinite 
thermal conductivity once again in the superfluid region.

\begin{figure}
\includegraphics[width=\linewidth,clip]{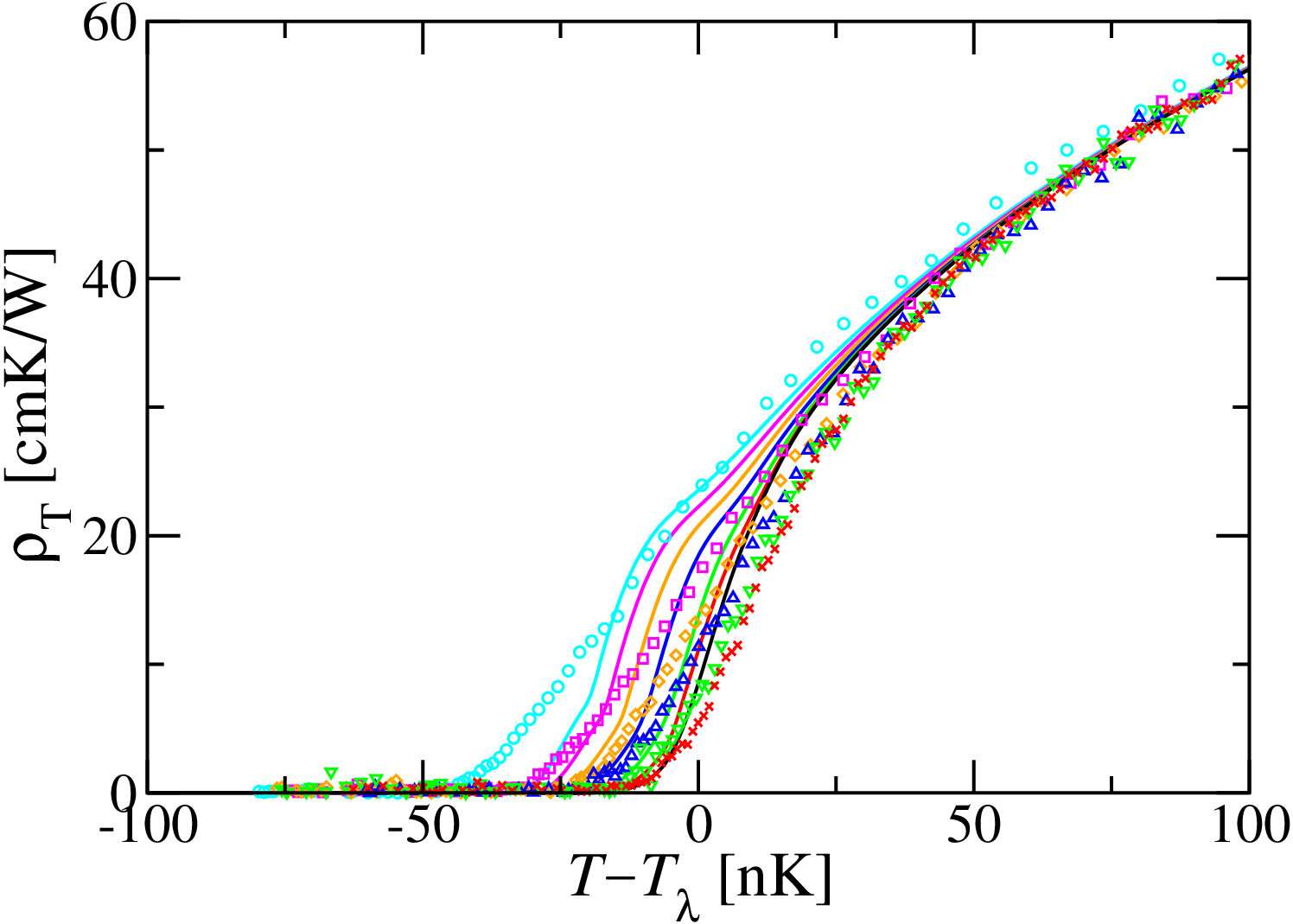}
\caption{(Color online) The thermal resistivity as a function of the 
temperature difference $T-T_\lambda$ for the superfluid/normal-fluid interface 
in gravity. The colored solid lines from left to right represent the heat currents 
$Q=160$, $130$, $100$, $70$, $40$, $20$, $0\,\mathrm{nW/cm^2}$. The data points 
represent the experimental data by Day \emph{et al.}\ \cite{Du98} taken for the 
same heat currents except $Q=0$. Related to the theoretical curves the data are 
ordered from left to right.}
\label{Fig:07}
\end{figure}

A related quantity is the thermal resistivity $\rho_\mathrm{T}=1/\lambda_\mathrm{T}$ 
which is the inverse of the thermal conductivity. In the normal-fluid region the 
thermal resistivity $\rho_\mathrm{T}$ is finite. On the other hand, in the superfluid 
region it is zero. For this reason the thermal resistivity $\rho_\mathrm{T}$ is well 
suited for a graphical representation. In Fig.\ \ref{Fig:07} the results of our 
numerical calculation are shown as colored solid lines for several positive values 
of the heat current $Q$. These heat currents are the same as those in the previous 
figures. Clearly, in the interface region near criticality the colored solid curves 
are smooth lines. The critical singularity at $T=T_\lambda$ is smoothed by the nonzero 
values of gravity $g$ and the heat current $Q$. For increasing heat currents $Q$ the 
colored solid lines are shifted to lower temperatures. This fact is again related 
to the depression of the superfluid transition temperature by nonzero heat currents 
following Duncan, Ahlers, and Steinberg \cite{DAS88}.

On both sides far away from criticality $T=T_\lambda$ and from the interface the 
colored curves asymptotically approach single lines, respectively. In the normal-fluid 
region the asymptotic thermal resistivity is given by \eqref{Eq:D11}, where in the 
superfluid region it is just zero.

The local thermal resistivity $\rho_\mathrm{T}$ has been measured for the 
superfluid/normal-fluid interface in the experiment by Day \emph{et al.}\ \cite{Du98}. 
While gravity on earth is $g=9.81\,\mathrm{m/s^2}$, the heat current is flowing 
upward from bottom to top so that $Q$ is positive. The values of heat current $Q$ 
are the same as in our calculation. For this reason, our colored solid lines can 
be compared directly with the experimental data. In Fig.\ \ref{Fig:07} the experimental 
data for several heat currents $Q$ are represented by points of several symbol types. 
The data points can be related to the solid curves by their color or alternatively 
by there order from left to right. An exception is the black solid line for zero 
heat current $Q=0$. In this latter case no experimental data are available. However, 
the black solid line is very close to the red solid line for $Q=20\,\mathrm{nW/cm^2}$. 
Consequently for $Q\lesssim 20\,\mathrm{nW/cm^2}$ the effect of the heat current is 
very small.

Qualitatively, the experimental data agree with the theoretical solid lines. However,
there are quantitative discrepancies. First of all, in the normal-fluid region well 
above criticality for temperature differences $T-T_\lambda\gtrsim 50\,\mathrm{nK}$ 
the thermal resistivity converges to a single line represented by the formula 
\eqref{Eq:D11}. We have multiplied the theoretical results with a correction factor 
which slightly differs from unity. In this way we achieve that far away from 
criticality the experimental data are lying on top of the theoretical curves. This 
correction factor is justified, because in Ref.\ \onlinecite{Do85} the model-$F$ 
parameters were adjusted for the theoretical specific heat and thermal resistivity 
in \emph{two-loop} order where in the present paper and also in our previous papers 
\cite{Ha99,Ha99a} the quantities were calculated in the Hartree approximation which 
is a self-consistent \emph{one-loop} approximation.

For larger heat currents $Q$ and temperatures $T$ slightly below $T_\lambda$ our 
solid lines show some bumps which are probably artifacts of our approximation. 
The magnitude of the artifacts is within the accuracy of our approach.
In the superfluid region well below criticality for temperature differences 
$T-T_\lambda\lesssim -50\,\mathrm{nK}$ the thermal resistivity $\rho_\mathrm{T}$ 
is very close to zero both in theory and experiment. This fact represents the 
frictionless heat transport by the superfluid/normal-fluid counterflow in 
superfluid $^4$He. Vortices are not present in our calculation presented here.

\subsection{Second sound and time dependent phenomena}
\label{Sec:04G}
Until now we have considered only stationary nonequilibrium states where a 
constant heat current $\mathbf{Q}=Q\,\mathbf{e}_z$ is flowing vertically in the 
$^4$He and where all time-dependent phenomena are relaxed. However, our numerical 
calculation solves the time-dependent model-$F$ equations so that time-dependent 
phenomena can be treated explicitly. In this case the order parameter 
$\langle\psi(z,t)\rangle$, the temperature $T(z,t)$, and the heat current density 
$\mathbf{Q}(z,t)=Q(z,t)\,\mathbf{e}_z$ are functions of altitude $z$ and time $t$. 
In superfluid $^4$He the most important dynamic phenomenon is second sound. We can 
generate a second sound pulse on the upper cell boundary $z_2$ if in the external 
heat source \eqref{Eq:D01} we choose a time-dependent upper source function 
$Q_2(t)$. Then the phase of the order parameter $\varphi(z,t)$, the temperature 
$T(z,t)$, and the local heat current $Q(z,t)$ show a pulse which is traveling 
downward toward the superfluid/normal-fluid interface. Approaching gradually 
the interface the width of the pulse increases. Once the interface is reached 
the pulse disappears by broadening where nearly nothing is reflected. In the end 
the second-sound pulse is absorbed nearly completely by the interface.

A heat pulse can be generated also on the lower cell boundary $z_1$ if the lower 
source function $Q_1(t)$ is chosen time dependent. However, in this case the 
response of the system is less spectacular because on the normal-fluid side 
the heat is transported diffusively. Nevertheless, our time-dependent investigations 
yield an important result. All time-dependent perturbations of the stationary 
states with a constant heat flow relax and disappear. Thus we conclude that the 
superfluid/normal-fluid interface in gravity and in the presence of a vertical heat 
flow is a stable physical configuration.

\section{Comparison with our previous approach for large heat currents}
\label{Sec:05}

\subsection{Solutions}
\label{Sec:05A}
We have found two different solutions of the model-$F$ equations for superfluid 
$^4$He in the nonequilibrium state where a heat current is flowing. In the first 
case the order parameter $\langle\psi\rangle=M\, e^{i\varphi}$ is nonzero 
and no vortices are present. The heat is transported convectively without any 
friction by the superfluid/normal-fluid counterflow so that the temperature 
gradient is zero. This solution is investigated in the present paper. In the 
second case the order parameter is zero due to fluctuations of the phase $\varphi$ 
by moving vortices and quantum turbulence. Here the moving vortices imply a small 
thermal resistivity which causes a small temperature gradient. This latter solution 
was investigated in our previous papers \cite{Ha99,Ha99a}. The two different 
solutions exist only in the superfluid and interface region where a nonzero order 
parameter is possible. In the normal-fluid region the solution is unique because 
here the order parameter is always zero.

The two solutions are controlled by the parameter $\sigma$ defined in 
Eq.\ \eqref{Eq:C85}. The first solution exists for negative and positive values 
of $\sigma$ where the second solution exists only for positive $\sigma$. For the 
superfluid/normal-fluid interface in gravity on earth this means that the first 
solution exist for small and large heat currents. On the other hand the second 
solution exists only for large heat currents $|Q| \gtrsim 70\,\mathrm{nW/cm^2}$ 
where the heat current is the major and gravity is the minor influence. However, 
the numerical calculations of our present paper show that in practice the first 
solution is stable only for small heat currents in the interval 
$-70\,\mathrm{nW/cm^2} \lesssim Q \lesssim +160\,\mathrm{nW/cm^2}$. For larger 
heat currents outside the interval we do not obtain a stable solution. We do not 
know if this instability is a property of our iteration procedure only which we 
describe in Subsec.\ \ref{Sec:03F}. However we guess that there is a physical 
instability beyond a certain critical heat current. This means we expect a 
discontinuous first-order like transition in the nonequilibrium state near the
heat current $Q_0\approx 70\,\mathrm{nW/cm^2}$ which separates the gravity dominated 
regime (first solution) from the heat current dominated regime (second solution). 
Nevertheless, there will be an overlap region where both solutions exist.

\subsection{Stability}
\label{Sec:05B}
The time-dependent nature of our numerical calculation provides a test for the 
physical stability of our stationary solutions. We may add a small perturbation 
to the solution and then start the calculation. After a time difference 
$\delta t\approx 2\,\mathrm{s}$ the perturbations relax and disappear so that the 
system returns to the stationary state. We find this behavior for all stationary 
solutions with a constant heat flow which we have presented in this paper. Thus, 
we conclude that the \emph{first-type solutions are stable} which describe 
physical states with a nonzero order parameter and no vortices. This result 
is expected because the order-parameter variation and second sound are damped. 
We note that we have a numerical instability for larger heat currents. However, 
this latter instability is unphysical and has a completely different nature 
because it appears on the short length scales $\Delta z$ of the discretization of 
the altitude coordinate.

In our previous papers \cite{Ha99,Ha99a} we did not prove the stability of the 
second-type solutions which describe superfluid states with vortices and a zero 
average order parameter. The reason is that in our previous papers we did not solve 
the model-$F$ equations as partial differential equations. Consequently, we do this 
now and provide the proof in the following. However, we note that in this case 
the stability is nontrivial because the system is in a superfluid state where the 
average order parameter is $\langle\psi\rangle=0$ and the temperature is 
$T<T_\lambda$. We solve the time-dependent renormalized model-$F$ equations for a 
second-type solution with a small perturbation. For simplicity we consider a 
self-organized critical state with a constant heat $Q$ current and constant 
temperature gradients $\nabla T=\nabla T_\lambda$ because this state is spatially 
homogeneous before the perturbation is applied. Unfortunately, the state is not 
periodic because we cannot require periodic boundary conditions for all quantities.  
Exceptions are the temperatures $T(z.t)$, $T_\lambda(z)$ and the phase of the order 
parameter $\varphi(z,t)$. Nevertheless, we can generalize the boundary conditions. For 
the latter three quantities we require pseudoperiodic boundary conditions in the sense 
of the \emph{impossible objects} of the famous Dutch graphic artist M.C. Escher 
\cite{Ho79}, i.e.
\begin{eqnarray}
T(z+L,t) &=& T(z,t) + \Delta T_L \ ,
\label{Eq:E01} \\
T_\lambda(z+L) &=& T_\lambda(z) + \Delta T_L \ ,
\label{Eq:E02} \\
\varphi(z+L,t) &=& \varphi(z,t) + \Delta\varphi_L(t) \ .
\label{Eq:E03}
\end{eqnarray}
Since the critical temperature $T_\lambda(z)$ is linear in $z$, from 
Eq.\ \eqref{Eq:E02} we obtain $\Delta T_L=(\partial_z T_\lambda) L$. On the other 
hand from Eq.\ \eqref{Eq:E01} we obtain $\Delta T_L=-(Q/\lambda_\mathrm{T})L$ where 
$\lambda_\mathrm{T}$ is the thermal conductivity. We note that for the self-organized 
critical state both results for the period constant $\Delta T_L$ must be equal. 
This fact implies $Q=-\lambda_\mathrm{T}(\partial_z T_\lambda)$.
The period constant of the order-parameter phase $\Delta\varphi_L(t)$ can be 
determined by investigating the renormalized model-$F$ equation \eqref{Eq:C79}. On the 
right-hand side we replace $\Delta\rho(z,t)$ in favor of the temperature $T(z,t)$ by 
inserting \eqref{Eq:C27}. We insert the complex order parameter $Y=\eta\,e^{i\varphi}$ 
and derive an equation for the order parameter phase $\varphi(z,t)$. We consider 
this equation for the altitudes $z=z_0$ and $z=z_0+L$ and then subtract the resulting 
equations. Thus, as a result we obtain 
$\partial_t \Delta\varphi_L(t) = g_0 (\Delta T_L/T_\lambda)$. We integrate this 
equation and obtain the period constant
\begin{equation}
\Delta\varphi_L(t) = \Delta\varphi_L(t_0) + g_0 (\Delta T_L/T_\lambda) (t-t_0) \ .
\label{Eq:E04}
\end{equation}
Clearly, the period of the order-parameter phase depends linearly on the time $t$. We 
conclude and find that for the self-organized critical state with small perturbations 
the renormalized model-$F$ equations can be solved numerically using Escher pseudoperiodic 
boundary conditions which are defined by Eqs.\ \eqref{Eq:E01}-\eqref{Eq:E03}. Our 
numerical test provides the following result. Any small perturbation of the 
self-organized critical state relaxes and disappears after a time difference of 
$\delta t\approx 2\,\mathrm{s}$. Thus we conclude that the \emph{second-type solutions 
are stable}. This means that the nonequilibrium states considered in the previous 
paper are stable.

A matter of special interest is the relaxation of the order parameter. We start 
at time $t_0$ with a small constant perturbation $Y(z,t_0)=\eta_0\,e^{i \varphi_0}$.
This means $\eta_0$ is small but nonzero, and $\varphi_0$ is constant. We consider 
the renormalized model-$F$ equation \eqref{Eq:C79} which describes the time evolution 
of the order parameter $Y(z,t)=\eta(z,t)\,e^{i\varphi(z,t)}$. Decomposing the 
equation with respect to the modulus $\eta(z,t)$ and the phase $\varphi(z,t)$ 
from the first term on the right hand side we infer the damping for the modulus
\begin{equation}
D= g_0 (2\gamma\,\tau) (w^\prime/F) [ \rho_1 + (\xi\nabla\varphi)^2 ] \ .
\label{Eq:E05}
\end{equation}
This damping is an inverse relaxation time. The solution of the model-$F$ equation 
is stable whenever this damping is positive and unstable otherwise. The prefactors 
are always positive so that the crucial quantity is the expression in the 
square brackets. For times $t$ shortly after the beginning of the calculation $t_0$ 
the phase is expected to be $\varphi(z,t)\approx \varphi_0$ so that 
$\nabla\varphi\approx \mathbf{0}$. Consequently, the main contribution is the 
dimensionless modified temperature parameter $\rho_1$. The stability of the 
solution depends on its sign. In Fig.\ \ref{Fig:01} $\rho_1$ is plotted as the 
blue dashed line for the superfluid/normal-fluid interface. A similar curve is 
obtained for the self-organized critical state if $\rho_1$ is plotted as a 
function of the heat current $Q$. While in the normal-fluid region $\rho_1$ 
is positive, in the superfluid region it is negative. For the heat current 
$Q=170\,\mathrm{nW/cm^2}$ we find the minimum value $\rho_{1,\mathrm{min}}=-0.39$. 
Thus, we conclude that the solution is stable in the normal-fluid region but 
unstable in the superfluid region.

However, the instability is true only for short times where $t-t_0$ is small. 
For longer times we must investigate the space and time dependence of the 
order-parameter phase $\varphi(z,t)$. From Eq.\ \eqref{Eq:E04} and from the 
renormalized model-$F$ equation \eqref{Eq:C79} we infer
\begin{eqnarray}
\varphi(z,t) &=& \varphi_0 + \Delta\varphi_L(t)\,(z-z_0)/L \nonumber\\
&=& \varphi_0 + g_0\,(\Delta T_L/T_\lambda)\,(t-t_0)\,(z-z_0)/L \ . \qquad
\label{Eq:E06}
\end{eqnarray}
Here $z_0$ is the altitude where the temperature equals the reference temperature, 
i.e. $T(z_0,t_0)=T_0$. For the last equality sign we have used \eqref{Eq:E04} 
realizing that in our case at the beginning $t=t_0$ the Escher period constant 
is $\Delta\varphi_L(t_0)=0$. Differentiating with respect to the altitude 
coordinate $z$ and multiplying by $\xi$ we obtain the dimensionless gradient of 
the order-parameter phase
\begin{equation}
\xi\nabla\varphi = g_0\,(\Delta T_L/T_\lambda)\,(t-t_0)\,(\xi/L)\,\mathbf{e}_z \ .
\label{Eq:E07}
\end{equation}
Consequently, we find $(\xi\nabla\varphi)^2 \sim (t-t_0)^2$. This means that in 
Eq.\ \eqref{Eq:E05} the second term in the square bracket increases with time. 
Even though at the beginning the square bracket may be negative because of the 
first term, after a short time the second term makes the square bracket positive. 
Thus we conclude: Even though there may be an instability for short times, the 
increase of the gradient of the phase \eqref{Eq:E07} makes the damping 
\eqref{Eq:E05} finally positive so that the time evolution of the order parameter 
is finally stable.

The stability of the second model-$F$ equation \eqref{Eq:C80} is easily proven. 
We insert the dimensionless renormalized heat current \eqref{Eq:C82} and 
neglect its last term because it is squared in the small nonzero order parameter. 
As a result we obtain a diffusion equation for the dimensionless renormalized 
temperature parameter $\Delta\rho(z,t)$. Since the related diffusion constant 
is positive, this equation is always stable. The covariant derivatives in this 
equation do not affect the stability.

We summarize that we have presented an explicit proof for the stability of the 
self-organized critical state which is a spatially homogeneous second-type solution 
of the model-$F$ equations. We expect that also the more general second-type 
solutions for spatially inhomogeneous systems are stable, which describe the 
superfluid/normal-fluid interface in our previous papers \cite{Ha99,Ha99a}. Our 
numerical calculations of the present paper support this expectation. However, 
we note that the stability is nontrivial because the system is in a superfluid 
state where the average order parameter is $\langle\psi\rangle=0$ and the 
temperature is $T<T_\lambda$.

\section{Discussion and conclusion}
\label{Sec:06}
Onuki \cite{On83,On87} and and later Weichman and Miller \cite{WM00} have 
also investigated the superfluid/normal-fluid interface within model $F$. They 
obtain temperature profiles which agree qualitatively with our results shown 
in Fig.\ \ref{Fig:03}. However, they did not use the renormalization-group 
theory and the related coupling parameters which have been determined by Dohm 
\cite{Do91}. For this reason, it is not possible to compare the results 
quantitatively. Weichman and Miller \cite{WM00} furthermore considered the 
self-organized critical state for a heat current flowing downward. While 
in the superfluid region usually the temperature profile is flat, they obtain 
phase slips in the order parameter which produce a stair-case like temperature 
profile. In this way they obtain a temperature gradient $\nabla T$ which on 
average equals the gradient $\nabla T_\lambda$ as required for the self-organized 
critical state. More recently Yabunaka and Onuki \cite{On10} performed a 
three-dimensional numerical simulation based on model $F$ in order to investigate 
the self-organized critical state and the superfluid/normal-fluid interface. 
They observed the formation and motion of vortices and phase slips which produce 
a nonzero temperature gradient $\nabla T$ on average in the superfluid region 
which compensates $\nabla T_\lambda$.

A sophisticated theory for mutual friction, quantum turbulence, and the dynamics 
of vortices in superfluid $^4$He was developed long time ago by Vinen \cite{Vi57}. 
A measure for the quantum turbulence is the density of the vortices which is 
defined as the total length of the vortex lines per volume. For this vortex density 
a rate equation is derived. On the right hand side of this equation there is a term 
for the generation and a term for the decay of vortices and quantum turbulence. 
Vortices are usually generated by a nucleation process. This means that an energy 
barrier must be overcome which strongly reduces the generation rate.

We believe that model-$F$ includes the effects of vortices and quantum turbulence 
correctly so that the Vinen theory can be derived if the model-$F$ equations can 
be solved exactly without any approximation. However, our two solutions which are 
derived within the Hartree approximation are idealized solutions of model $F$. 
They describe the two phases of a first-order nonequilibrium transition but do 
not include metastability and the nucleation process. 

For the check between theory and experiment an important quantity is the thermal 
resistivity $\rho_\mathrm{T}$ in superfluid $^4$He for $T<T_\lambda$ and larger 
heat currents $Q\gtrsim 100\,\mathrm{nW/cm^2}$ induced by the effect of vortices. 
In Fig.\ 3 of our previous paper \cite{Ha99a} we have compared the result of our 
second-type solution with experimental data by Baddar \emph{et al.}\ \cite{Al00}. 
The experimental thermal resistivity is lower by a factor of $20$ than our theoretical 
result. A plausible explanation for this discrepancy is the following. In the 
experiment the heat current is flowing upward from bottom to top. The related 
superfluid current is flowing in the opposite direction, i.e.\ downward from top 
to bottom. Consequently, the vortices are transported together with the superfluid 
current downward. Since the order parameter $\langle\psi\rangle$ increases and 
the vortex density decreases with the altitude $z$, the downward transport of the 
vortices together with the metastability of the nucleation process can reduce 
the vortex density considerably. Since the hopping over energy barriers in the 
nucleation process causes exponential factors, a reduction of the thermal 
resistivity by a factor $20$ is plausible. 

The situation is different for the self-organized critical state. Here the system 
is spatially homogeneous because the heat current $Q$ and the temperature difference 
$T(z)-T_\lambda(z)=\Delta T$ are constant and do not depend the altitude $z$. The 
nucleation process reaches an equilibrium state, so that our theory will predict 
the density of the vortices and the related thermal resistivity correctly. A first 
agreement with the theory was found in the experiment by Moeur \emph{et al.}\ 
\cite{Du97}. Here the temperature difference $\Delta T$ was measured as a function 
of the heat current $Q$. The comparison with our theory is shown in Fig.\ 3 of our 
previous paper \cite{Ha99}. The agreement is encouraging. However, for very large 
heat currents $Q\gtrsim 2\,\mathrm{\mu W/cm^2}$ deep in the superfluid region a 
deviation was found. This deviation may be a problem of the temperature measurement 
because the temperature $T$ is never measured in the bulk of the system but always 
on the surface. In order to avoid the Kapitza resistance which implies a temperature 
jump on the surface, the temperature is usually measured by thermometers on the side 
walls. However, in cases where vortices are present, a superfluid flow parallel 
along a side wall may cause a transverse Kapitza resistance so that there is a 
temperature jump also on a side wall. 

In a recent experiment Chatto \emph{et al.}\ \cite{Du07} performed an experiment 
to measure the thermal conductivity/resistivity indirectly where the explicit 
measurement of the temperature is avoided. Instead they measured the velocity $v$ 
of a propagating thermal mode as a function of the heat current $Q$ in the interval 
$30\,\mathrm{nW/cm^2}\lesssim Q\lesssim 15\,\mathrm{\mu W/cm^2}$. On the other 
hand, they derived a theoretical curve for the velocity $v$ from our theoretical 
results for the thermal conductivity \cite{Ha99,Ha99a}. They find very good agreement 
between theory and experiment in the whole range of heat currents $Q$, even for 
the largest values $Q\approx 15\,\mathrm{\mu W/cm^2}$ deep in the superfluid 
region. We conclude that this experiment is an important verification of our 
theory. This means our theory \cite{Ha99,Ha99a} describes the effects of vortices, 
mutual friction, and the thermal conductivity/resistivity correctly on a quantitative 
level for the self-organized critical state.

\acknowledgments
\noindent
We would like to thank Prof.\ Dr.\ V.\ Dohm for helpful comments on the manuscript 
and Prof.\ Dr.\ J.A. Lipa for experimental data of the specific heat.

\end{document}